\documentclass[aps,prx,reprint,superscriptaddress,letterpaper,twocolumn,longbibliography,floatfix]{revtex4-2}

\usepackage{graphicx}

\usepackage{amsmath}
\usepackage{mathtools}
\usepackage{amsfonts}
\usepackage{amssymb}
\usepackage{bm}

\newcommand{\bx}{\boldsymbol{x}}

\usepackage{listings}
\usepackage{xcolor}

\definecolor{dark-red}{rgb}{0.4,0.15,0.15}
\definecolor{dark-blue}{rgb}{0.15,0.15,0.4}
\definecolor{medium-blue}{rgb}{0,0,0.5}
\usepackage[hypertexnames=false]{hyperref}
\usepackage[all]{hypcap}
\hypersetup{
	colorlinks, linkcolor={dark-red},
	citecolor={dark-blue}, urlcolor={medium-blue}
}

\definecolor{bkgd}{RGB}{240,242,246}
\definecolor{ceruleanblue}{rgb}{0.16, 0.32, 0.75}
\definecolor{orange-red}{rgb}{1.0, 0.27, 0.0}
\definecolor{anotherblue}{RGB}{37,92,243}
\definecolor{blackblue}{RGB}{46,60,85}
\definecolor{goldyellow}{RGB}{199,146,12}
\lstdefinestyle{altstyle2}{
	backgroundcolor=\color{bkgd},
	basicstyle=\ttfamily\footnotesize\color{blackblue},
	breakatwhitespace=false,
	breaklines=true,
	captionpos=b,
	commentstyle=\color{goldyellow},
	keepspaces=true,
	keywordstyle=\color{orange-red},
	language=Python,
	numbersep=5pt,
	numberstyle=\tiny\color{ceruleanblue},
	showspaces=false,
	showstringspaces=false,
	showtabs=false,
	stringstyle=\color{anotherblue},
	tabsize=2
}
\usepackage{dsfont}

\begin{document}
\title{Improved diffusive approximation of Markov jump processes close to equilibrium}

\author{David Roberts}
\author{Trevor McCourt}
\thanks{Corresponding author: \href{mailto:trevor@extropic.ai}{trevor@extropic.ai}}
\author{Geremia Massarelli}
\author{Jeremy Rothschild}
\author{Nahuel Freitas}

\affiliation{Extropic Corporation, Cambridge, Massachusetts, USA}

\date{\today}
\begin{abstract}
    Diffusive approximations of Markov jump processes often fail to accurately capture large fluctuations. This is confounding, as the rare events triggered by these large fluctuations, such as the failure of electronic memories, are often the object of interest. In this paper we present an improved diffusive approximation, extending a method previously limited to equilibrium systems. Using new tools from stochastic thermodynamics, we prove its validity to linear order in departures from equilibrium and demonstrate its superior accuracy over the Kramers-Moyal expansion in predicting both steady-state and transient properties, including the error rate of a non-equilibrium electronic memory.
\end{abstract}
\maketitle

\section{Introduction}

Markov jump processes (MJPs) in continuous time are ubiquitous across many different disciplines, being employed to model stochastic phenomena in biology, chemistry, physics and finance. For example, continuous-time MJPs are the basis for the stochastic description of elementary chemical reactions in cells, and also for effective models of their gene regulatory networks \cite{Arkin1998, Bressloff2017}. In physics, MJPs are at the core of important models of non-equilibrium systems like lattice-gas models \cite{Spohn1983, Lebowitz1999} and, among other examples, are employed to describe the transport of electrons across both nanoscopic \cite{Wasshuber, Moreira2023} and mesoscopic \cite{Gao2021, freitasStochastic2021, freitas2025} electronic circuits.

While exact algorithms to simulate MJPs exist~\cite{gillespie2976general, gillespie1977exact}, it is often convenient to approximate MJPs by diffusive processes typically described by Langevin dynamics at the trajectory level or Fokker–Planck equations at the ensemble level. A continuous representation is often advantageous---analytically (e.g., closed--form asymptotics, perturbation, spectral methods) and computationally. Examples in which diffusion approximations are particularly effective include chemical Langevin equations \cite{gillespie2000chemical, Horowitz2015} in biochemistry or stochastic Lotka–Volterra models~\cite{mckane2005predator, fisher2014transition} in population dynamics.

\begin{figure}[ht!]
    \centering
    \includegraphics[width=0.95\linewidth]{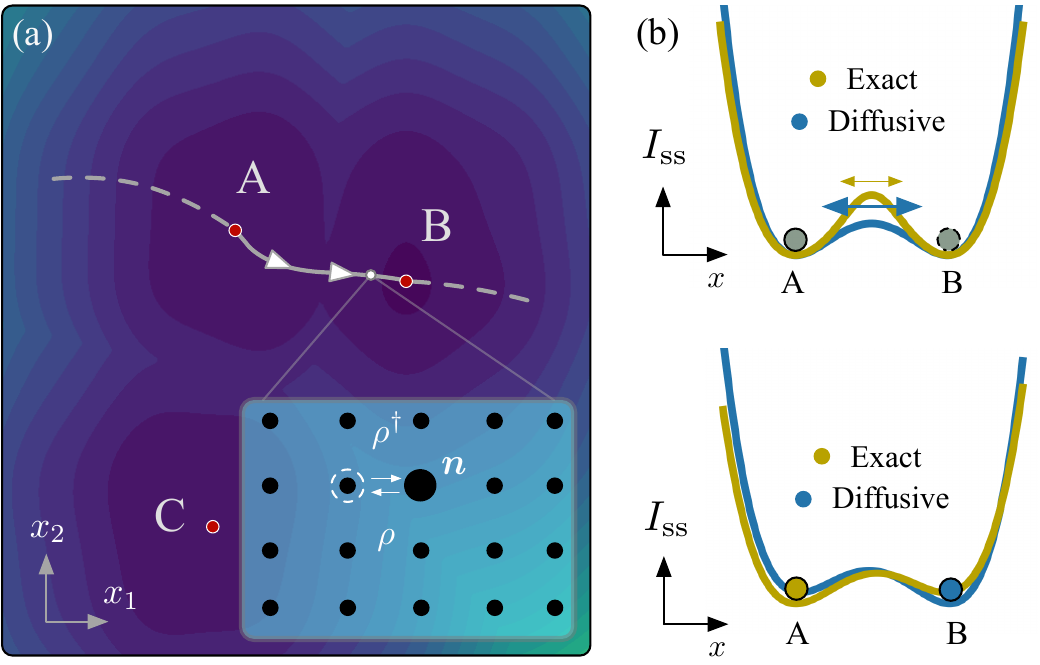}
    \caption{{\bf Diffusive approximation of Markov Jump Processes}. 
    (a) The steady-state probability distribution $P_\text{ss}$ for a two-dimensional stochastic process with three metastable states ($A$, $B$, and $C$). The white line indicates a transition path from $A$ to $B$, driven by stochastic fluctuations. Inset: A ``zoom-in'' on the transition path reveals the underlying discrete Markov jump process. (b) A comparison of the effective energy landscape along the path from $A$ to $B$, given by the steady-state self-information $I_\text{ss}:=-\log P_\text{ss}$ for the exact process (yellow) and a typical diffusive approximation (blue). The following panels highlight common failure modes of standard diffusive approximations that our improved method is designed to address. Top panel: The approximation underestimates the barrier height between the two metastable states. Bottom panel: While standard diffusive approximations correctly capture local behavior around stable states, their inaccuracy for large fluctuations can lead to qualitatively wrong predictions, such as identifying the wrong most-stable state (the global maximum of $P_\text{ss}$) \cite{vellela2009}.  }
    \label{fig:intro}
\end{figure}

Thus, the diffusive approximation of MJPs is an important problem in statistical physics with a long ``confused history'' \cite{Gardiner, gillespie2000chemical}, for which different methods have been developed. The truncation of the Kramers-Moyal expansion~\cite{Gardiner, reichl} and the van Kampen Linear Noise Approximation are the most popular~\cite{vankampen}.

However, it is well-known that popular diffusive approximations are only guaranteed to be accurate for small, typical fluctuations, and fail to capture the probability of large, rare fluctuations \cite{hanggi1982, hanggi1984, hanggi1988bistability}. This is relevant in problems displaying metastability, where rare fluctuations are involved in the switching events between different dynamical attractors. Prominent examples are given by extinction events in population dynamics models~\cite{reichenbach2006coexistence, ovaskainen2010stochastic,panahi2025}, metastable gene expression patterns in gene regulatory networks~\cite{Bressloff2017}, or the occurrence of thermally induced soft errors in bistable electronic circuits \cite{rezaei2020, freitasReliability2022}.

To address this limitation, we introduce an improved diffusive approximation for Markov jump processes that remains accurate for arbitrarily large fluctuations, even in non-equilibrium systems. Our central contribution is to generalize a method, previously restricted to systems at detailed balance, by proving that it correctly captures macroscopic fluctuations to linear order in the departure from equilibrium.

Our work provides a more robust and systematically improvable framework for studying rare events using diffusion approximations. We validate our approach on models of stochastic electronic circuits, demonstrating that it more accurately predicts both steady-state distributions and key transient properties, such as the error rate of a bistable memory, when compared to standard approximations. This improved accuracy is not merely a theoretical curiosity: the modeling framework for stochastic electronic circuits that is featured in the examples studied here has been successfully applied to the design and experimental validation of novel probabilistic circuits in the companion articles \cite{freitas2025, jelincic2025}.

The key to our improved approximation is a modification of the diffusion tensor in the Fokker-Planck equation. While the standard Kramers-Moyal approximation uses a simple sum of forward and reverse jump rates, our formulation employs their logarithmic mean. This seemingly minor change has profound consequences, ensuring the resulting diffusive process correctly captures large fluctuations in the steady state for systems that are close to being detailed-balanced. While primarily tailored to reproduce steady-state fluctuations, our method also yields more precise estimates for key dynamical properties. Crucially, by better capturing the behavior in the tails of higher-order eigenvectors of the generator of the jump process, our approximation provides a more faithful description of large fluctuations not just at steady state, but throughout the system's transient evolution.

This article is organized as follows. Section~\ref{sec:background} introduces both the standard diffusive approximation of Markov jump processes, namely the Kramers-Moyal approximation, as well as the unconventional diffusion approximation that we study in this work. Section \ref{sec:eff_FP} then presents its derivation. In Section \ref{sec:dynamical-ldt} we prove fundamental accuracy guarantees for the improved diffusion approximation using recent results in macroscopic stochastic thermodynamics. In Section \ref{sec:examples} we compare the performance of the usual Kramers-Moyal approximation and the improved approximation by applying them to several models of stochastic non-linear electronic circuits. Finally, we summarize our results in Section \ref{sec:discussion} and mention possible connections with recent work by other authors.

\section{Markov jump processes and diffusive approximations}
\label{sec:background}

In this section, we present our main theoretical result: an improved diffusive approximation for a general Markov jump process over states $\bm{n} \in \mathbb{Z}^d$ (see Figure \ref{fig:intro}). For example, each state can represent the copy-numbers of different species in chemical reaction networks \cite{Schmiedl2007, Lazarescu2019}, or the number of charges in each node of an electronic circuit \cite{freitasStochastic2021, Gao2021}. The state of the system evolves through transitions $\bm{n} \to \bm{n} + \bm{\Delta}_\rho$, which have associated jump vector $\bm{\Delta}_\rho$ and jump rate $\chi_\rho(\bm{n})$. Thus, transitions are indexed by $\rho \in \mathbb{Z}^d\setminus\{0\}$. Given any transition with $\rho>0$, the reverse transition is indexed by $\rho^\dagger = -\rho$ and has jump vector $\bm{\Delta}_{-\rho} = -\bm{\Delta}_\rho$. For some initial condition, let $P(\bm{n}, t)$ be the probability of the system being in state $\bm{n}$ at time $t$. This probability evolves according to the master equation
\begin{equation}
    \partial_t P(\bm{n},t) =\! 
    \sum_{\rho} \!\! \bigg[  \chi_\rho(\bm{n}-\bm{\Delta}_\rho) P(\bm{n}-\bm{\Delta}_\rho, t)
    -\chi_\rho(\bm{n})P(\bm{n},t) \! \bigg]\!.\label{eq:cme}
\end{equation}
In chemical reaction network theory, Eq.~\eqref{eq:cme} is the \emph{chemical master equation} (CME), the standard stochastic description in the low--copy--number regime where discreteness and intrinsic noise are essential. For systems in which the jump rates depend on some scale parameter $\Omega$, we can define the density $\bm{x} = \bm{n}/\Omega$ and 
the density-dependent rates $\lambda_\rho(\bm{x}) \equiv \chi_\rho(\Omega \bm{x})$~\cite{gillespie2000chemical, vankampen}. Then, defining the rescaled jump vectors $\bm{x}_\rho \equiv \bm{\Delta}_\rho/\Omega$, the master equation for the probability $P(\bm{x},t)$ is
\begin{equation}
\begin{split}
    \partial_t P(\bm{x},t) &=\!  
    \sum_{\rho} \!\bigg[ \! \lambda_\rho(\bm{x}\!-\!\bm{x}_\rho) P(\bm{x}\!-\!\bm{x}_\rho, t)-  \lambda_\rho(\bm{x})P(\bm{x},t) \! \bigg] \!.
    \label{eq:ME}
\end{split}
\end{equation}
\noindent At the ensemble level, general diffusive processes are described by Fokker-Planck equations of the form:
\begin{equation}
    \partial_t P(\bm{x},t) = -\partial_n \Big[ \mu_n(\bm{x}) P(\bm{x},t)-\partial_m(D_{nm}(\bm{x}) P(\bm{x},t))\Big],
    \label{eq:FP}
\end{equation}
in terms of a drift vector field $\mu_n(\bm{x})$ and a diffusion tensor field $D_{nm}(\bm{x})$. In the previous equation, Einstein summation is assumed and $\partial_n \equiv \partial/\partial{x_n}$. A generic method to obtain a diffusive approximation of a given MJP is given by the second order truncation of the Kramers-Moyal expansion~\cite[VIII.5]{vankampen}. According to this method, the drift and diffusion fields are constructed as:
\begin{equation}
    \mu_n(\bm{x}) \equiv \frac{1}{\Omega}\sum_{\rho>0} \left( \lambda_\rho(\bm{x}) - \lambda_{-\rho}(\bm{x}) \right) \Delta_\rho^n,
    \label{eq:drift_1}
\end{equation}
and 
\begin{equation}
    D_{nm}^\text{KM}(\bm{x}) \equiv \frac{1}{2\Omega^2} \sum_{\rho>0} 
    \left( \lambda_\rho(\bm{x}) + \lambda_{-\rho}(\bm{x}) \right)
    \Delta_\rho^n \Delta_\rho^m,
    \label{eq:kramers_moyal_diff_1}
\end{equation}
where $\Delta_\rho^n$ is the $n$-th component of $\bm{\Delta}_\rho$.
As mentioned above, this approach fails to capture large fluctuations accurately. To make this statement more precise, let us define $P_\text{ss}^\text{ME}(\bm{x})$ as the steady-state distribution according to the master equation in Eq.~\eqref{eq:ME}, and $P_\text{ss}^\text{KM}(\bm{x})$ as the steady-state distribution according to Eqs.~\eqref{eq:FP}, \eqref{eq:drift_1} and \eqref{eq:kramers_moyal_diff_1}. Under the assumption that the transition rates $\lambda_\rho(\bm{x})$ are extensive, it can be seen that both $P_\text{ss}^\text{ME}(\bm{x})$ and $P_\text{ss}^\text{KM}(\bm{x})$ decrease exponentially in the limit $\Omega \to \infty$. This means that the \emph{rate functions} $I_\text{ss}^\text{ME/KM}(\bm{x}) \equiv \lim_{\Omega \to \infty}-\log(P_\text{ss}^\text{ME/KM}\left(\bm{x})\right)/\Omega
$ are well-defined and nonnegative. Furthermore, it can be shown~\cite{hanggi1988bistability} that the functions $I_\text{ss}^\text{ME}(\bm{x})$ and $I_\text{ss}^\text{KM}(\bm{x})$ share the same local minima (the most probable values of $\bm{x}$) and actually match each other to second-order around those minima (small Gaussian fluctuations, see Fig.~\ref{fig:intro}, top panel). However, it is also possible to see that 
$I_\text{ss}^\text{ME}(\bm{x}) \neq I_\text{ss}^\text{KM}(\bm{x})$ in general, which means that the approximation given by Eqs.~\eqref{eq:drift_1} and \eqref{eq:kramers_moyal_diff_1} fails to capture the probability of large steady-state fluctuations, \emph{even to dominant order in the macroscopic limit}. In Fig.~\ref{fig:intro}b we illustrate an example in which the height of the barrier separating two minima of $I_\text{ss}(\bm{x})$, which is related to the rate of switching transitions, is underestimated by the Kramers–Moyal approximation. 
More seriously, even if both functions share the same local minima, it is known that their global minima can be different, which leads to incompatible predictions in the macroscopic limit (see the bottom panel of Fig.~\ref{fig:intro}b).

It is useful to contrast our setting with two related lines of work. First, in studies of stochastic transport with tilted periodic potentials (e.g., molecular motors), the underlying dynamics is taken to be continuous (Smoluchowski/Langevin), and discrete Kramers–type descriptions are used as surrogates \cite{alamilla2020, challis2016,hayashi2015}. Our problem is the complementary direction: given an MJP, we construct a diffusion that preserves its large–deviation structure (exactly at equilibrium and to first order out of equilibrium). Second, there is a body of work starting from generalized (often non-Markovian) Langevin dynamics that introduces effective temperatures or chemical potentials to extend fluctuation–dissipation relations far from equilibrium \cite{holek2020}. By contrast, we start from a discrete, locally detailed-balanced MJP and, in the large-system-size limit, derive a modified diffusion tensor that preserves its large–deviation
structure (exactly at equilibrium and to first order out of
equilibrium).

The main result in this article is that, for a large class of MJPs, one must instead consider the following diffusion tensor \footnote{It is useful to point out that, provided that the jump vectors $\{\mathbf{\Delta}_\rho\}_\rho$ are linearly independent, then near a deterministic fixed point $\bm{x}\approx \bm{x}_*$, the logarithmic mean appearing in Eq. \eqref{eq:improved_diff} reduces to the arithmetic mean, so the above diffusion tensor and Eq.~\eqref{eq:kramers_moyal_diff_1} asymptotically coincide near a deterministic fixed point.}:
\begin{equation}
    D_{nm} (\bm{x}) \equiv \frac{1}{\Omega^2}\sum_{\rho>0} 
    \frac{\lambda_\rho(\bm{x}) - \lambda_{-\rho}(\bm{x})}{\log \left(\lambda_\rho(\bm{x})/\lambda_{-\rho}(\bm{x})\right)}
    \Delta_\rho^n \Delta_\rho^m.
    \label{eq:improved_diff}
\end{equation}
Equation~\eqref{eq:improved_diff} is the generalization to arbitrary MJPs of a result first obtained in \cite{hanggi1982, hanggi1984, hanggi1988bistability} for one-dimensional systems with nearest-neighbor jumps. There, it was also shown that the resulting diffusive dynamics has exactly the same macroscopic steady-state fluctuations as the original MJP under strict detailed balance (which is always satisfied in such 1D systems). In this article, we leverage a connection between the rate function of NESSs and the entropy produced along deterministic trajectories \cite{freitas2021linear, freitas2022emergent, falasco2023, santolin2024} to prove that Eq.~\eqref{eq:improved_diff} also reproduces steady-state macroscopic fluctuations to linear order in departures from strict detailed balance. Finally, we also note that the quantity $\frac{\lambda_\rho(\bm{x}) - \lambda_{-\rho}(\bm{x})}{\log \left(\lambda_\rho(\bm{x})/\lambda_{-\rho}(\bm{x})\right)}$ that enters in Eq.~\eqref{eq:improved_diff} has recently been identified in Refs.~\onlinecite{VanVu2023, VanVu2024} to play a crucial role in the connection between stochastic thermodynamics and optimal transport theory. As explained in those works, it can be interpreted as a microscopic and non-linear version of the usual Onsager response coefficients.

The theory behind Eq.~\eqref{eq:improved_diff} is tailored to accurately reproduce steady-state fluctuations, i.e., to improve the correspondence between the zero modes of the generators of the master equation in Eq.~\eqref{eq:ME} and the Fokker–Planck dynamics in Eq.~\eqref{eq:FP}. However, as we show, it also improves the correspondence between higher-order eigenmodes and eigenvalues of both dynamics, compared to the regular diffusive approximation based on Eq.~\eqref{eq:kramers_moyal_diff_1}. In particular, we show that our theory improves the estimation of the error rate of a bistable electronic memory, which is given by the first nonzero eigenvalue of the MJP generator \cite{freitasReliability2022}.

\section{Derivation of the modified diffusion approximation}
\label{sec:eff_FP}

\subsection{Effective Fokker-Planck equation}
In order to derive Eq.~\eqref{eq:improved_diff}, our strategy will be to construct a drift vector $\mu_n(\bm{x})$ and a diffusion tensor $D_{nm}(\bm{x})$ such that the dynamics in Eq.~\eqref{eq:FP} mimics that of Eq.~\eqref{eq:ME}.
The first step in our derivation is to recast the master equation in Eq.~\eqref{eq:ME} in the form of a Fokker-Planck equation, but for which the diffusion tensor depends on the probability distribution itself. Indeed, with some manipulation of Eq.~\eqref{eq:ME} we obtain the exact equation of motion
\begin{align}
    &\partial_t {P(\bm{x},t)} = \sum_\rho  \Bigg[\sum_{k=1}^\infty \frac{1}{k!}\bigg(\frac{-\bm{\Delta}_\rho\cdot \nabla}{\Omega}\bigg)^k (\lambda_\rho P)\Bigg]\label{eq:non-linear-FP}\\
    &=-\partial_n \big[\mu_n(\bm{x}) P(\bm{x},t)  - \partial_m \big(\widetilde{D}_{nm}(\bm{x},t) P(\bm{x},t)\big) \big]\label{eq:non-linear-FP2}
\end{align}
where (see \footnote{
To see this, note that, defining $\bm{x}_\rho \equiv \boldsymbol{\Delta}_\rho/\Omega$, we can write
\[\partial_t P = \sum_\rho  \Bigg(\partial_n x_\rho^n (\lambda_\rho P)~~~~~~~~~~~~~~~~~~~~~~~~~~~~~~~~~~\]
\[~~~~~~~~~~~~~~~~~+ \partial_n \partial_m x_\rho^m x_\rho^n\sum_{k=2}^\infty \frac{(\partial_n x_\rho^n)^{k-2}}{k!} (\lambda_\rho P)\Bigg)\]
We then note that the above is in the form Eq. \eqref{eq:non-linear-FP2}, with $\mu_n(\bm{x}) = \sum_\rho x_\rho^n \lambda_\rho(\bm{x})$ and $\tilde{D}(\bm{x},t)$ given by Eq. \eqref{eq:non-linear-diff}.} for a brief derivation)
\begin{align}
    \mu_n(\bm{x})& \equiv \frac{1}{\Omega}\sum_\rho \lambda_\rho(\bm{x}) \Delta_\rho^n
    \label{eq:drift_2}\\
    \widetilde D_{nm}(\bm{x},t)& \equiv  
    \frac{1}{\Omega^2}\sum_\rho F_\rho(\bm{x},t) \Delta_\rho^n \Delta_\rho^m,
\label{eq:non-linear-diff}
\end{align}
with
\begin{equation}
    F_\rho(\bm{x},t)  \equiv
    e^{\mathcal I(\bm{x},t)}\sum_{k \geq 2} \frac{1}{k!}\bigg(\frac{-\bm{\Delta}_\rho\cdot \nabla}{\Omega}\bigg)^{k-2}\left[\lambda_\rho(\bm{x}) e^{-\mathcal I(\bm{x},t)}\right].
\label{eq:non-linear-diff-core}
\end{equation}
In the previous equation, we have introduced the self-information or surprisal associated to the distribution $P(\bm{x},t)$:
\begin{equation} \label{eq:self_information}
    \mathcal{I}(\bm{x},t) \equiv -\log(P(\bm{x},t)).
\end{equation}
Note that Eq.~\eqref{eq:non-linear-FP} cannot be interpreted as a regular Fokker-Planck equation since the function $\tilde D_{nm}(\bm{x})$ depends implicitly on the probability distribution itself.
We now discuss different approximations that can be performed on Eq.~\eqref{eq:non-linear-diff-core} in order to reduce Eq.~\eqref{eq:non-linear-FP} to a regular Fokker-Planck dynamics.

\emph{Kramers-Moyal truncation.} In the first place, we notice that if only the $k=2$ term in the summation of Eq.~\eqref{eq:non-linear-diff-core} is kept, then the diffusion tensor $\widetilde{D}_{nm}(\bm{x})$ becomes independent of $P(\bm{x},t)$ and reduces to:
\begin{equation}
    D_{nm}^\text{KM}(\bm{x}) \equiv \frac{1}{2}\sum_\rho \lambda_\rho(\bm{x})\frac{\Delta_\rho^n \Delta_\rho^m}{\Omega^2}
    \label{eq:kramers_moyal_diff_3}
\end{equation}
In this case, Eq.~\eqref{eq:non-linear-FP} coincides with the Fokker-Planck equation obtained by the usual truncation to second order of the Kramers-Moyal expansion~\cite{vankampen, risken}. As mentioned before, this method only captures small fluctuations around the most probable values of $\bm{x}$.

\emph{Steady-state matching.} We now consider situations in which the master equation in Eq.~\eqref{eq:ME} has a unique steady state $P_\text{ss}(\bm{x})$. Then, one possible approximation is to replace $\mathcal{I}(\bm{x},t)$ in Eq.~\eqref{eq:non-linear-diff-core} with the steady-state self-information $\mathcal{I}_\text{ss}(\bm{x}) = -\log(P_\text{ss}(\bm{x}))$, obtaining:
\begin{equation}
\begin{split}
    F_\rho^\text{SS}(\bm{x}) \equiv  e^{\mathcal{I}_\text{ss}(\bm{x})}
    &\sum_{k \geq 2} \frac{1}{k!} \left(\frac{-\bm{\Delta}_\rho \cdot \nabla}{\Omega}\right)^{k-2}\left[\lambda_\rho(\bm{x}) e^{-\mathcal{I}_\text{ss}(\bm{x})}\right].
\end{split}
\label{eq:ss_match_diff}
\end{equation}
We can then treat this approximation as a time-independent ansatz for the diffusion tensor, $\widetilde{D}(\bm{x},t)\equiv \widetilde{D}(\bm{x})$ appearing in Eq. \eqref{eq:non-linear-FP2}. The regular Fokker-Planck dynamics that results from this approximation is guaranteed to have exactly the same steady-state distribution as the original Markov jump process. However, this ansatz is hardly useful, since the steady-state distribution is unknown in most cases. Some exceptions are one-dimensional jump processes with nearest-neighbor transitions~\cite{hanggi1988bistability}, or the kind of detailed-balanced processes discussed next.

\emph{Strict detailed-balance.} If the transition rates $\lambda_{\rho}(\bm{x})$ satisfy the detailed-balance conditions,
\begin{equation}
    \log \frac{\lambda_\rho(\bm{x})}{\lambda_{-\rho}(\bm{x} + \bm{x}_\rho)} = - (\Phi(\bm{x}+\bm{x}_\rho) - \Phi(\bm{x})), 
    \label{eq:gdb}
\end{equation}
in terms of some state function $\Phi(\bm{x})$, then the steady state of the Markov jump process is given by the Gibbs equilibrium distribution
\begin{equation}
    P_\text{eq}(\bm{x}) = \frac{e^{-\Phi(\bm{x})}}{Z},
\end{equation}
with $Z = \sum_{\bm{x}}e^{-\Phi(\bm{x})}$. Therefore, 
$\mathcal{I}_\text{ss}(\bm{x}) = \Phi(\bm{x}) + \log(Z)$. Introducing this in Eq.~\eqref{eq:ss_match_diff} we obtain the detailed-balance result
\begin{equation}
    F_\rho^\text{DB}(\bm{x}) \equiv  F^\text{SS}(\bm{x}) |_{\mathcal{I}_\text{ss} = \Phi + \log(Z)}.
\end{equation}
Having constrained the parameters to yield identical steady-state distributions, we can now probe the deeper similarities and differences between the models. To this end, we turn in Sec. IV.A to the linear response framework, which provides the tools to analyze both the dynamical response and key statistical features of the steady state itself, including the self-information.

\subsection{Scaling limit}
\label{sec:scaling_limit}
The expression in Eq.~\eqref{eq:non-linear-diff-core} is exact. We now discuss how Eq.~\eqref{eq:non-linear-diff-core} is simplified by taking the limit $\Omega \to \infty$. We assume that the jump rates $\lambda_\rho$ are extensive quantities and expand $\lambda_\rho(\bm{x}) \equiv \Omega \omega_\rho(\bm{x}) + o(\Omega)$. In that case, it can be shown that the distribution $P(\bm{x},t)$ satisfies a large-deviation principle, which means that the self-information $\mathcal{I}(\bm{x},t)$ is also extensive  \cite{kubo1973fluctuation}. Thus, we consider the expansion:
\begin{align}
    \mathcal I(\bm{x},t) &= \Omega I(\bm{x},t) + o(\Omega) 
        \label{eq:self-information-expansion}    
\end{align}
One can then show that the function $F_\rho(\bm{x})$ involved in Eq.~\eqref{eq:non-linear-diff} admits an identical expansion:
\begin{equation} \label{eq:non-linear-diff-core-expansion}
   F_\rho(\bm{x}) = \Omega f_\rho(\bm{x}) + o(\Omega)
\end{equation}
As shown in Sec. III of the Supplemental Materials (SM), $f_\rho(\bm{x})$ then takes the form:
\begin{equation}
\begin{split}
    f_\rho(\bm{x}) &= 
    \omega_\rho(\bm{x}) 
    \sum_{k\geq 2} \frac{1}{k!} (\bm{\Delta}_\rho \cdot \nabla I(\bm{x},t))^{k-2}\\
    &= \omega_\rho(\bm{x})\: \frac{e^{\bm{\Delta}_\rho \cdot \nabla I(\bm{x},t)}-1-\bm{\Delta}_\rho \cdot \nabla I(\bm{x},t)}{(\bm{\Delta}_\rho \cdot \nabla I(\bm{x},t))^2}.
\end{split}
\label{eq:f_scaling}
\end{equation}
It is instructive to specialize the last expression for the case of equilibrium matching. For this, we expand $\Phi (\bm{x}) = \Omega \phi(\bm{x}) + o(\Omega)$. Then, in the scaling limit $\Omega\to\infty$, the detailed-balance conditions in Eq.~\eqref{eq:gdb} read as:
\begin{equation}
    \log \frac{\omega_\rho(\bm{x})}{\omega_{-\rho}(\bm{x})} = - \bm{\Delta}_\rho\cdot\nabla \phi(\bm{x}).\label{eq:macro-db}
\end{equation}
Therefore, we have:
\begin{equation}
\bm{\Delta}_\rho \cdot \nabla I_\text{ss}(\bm{x}) = \bm{\Delta}_\rho \cdot \nabla \phi(\bm{x}) = -\log \frac{\omega_\rho(\bm{x})}{\omega_{-\rho}(\bm{x})}   
\end{equation}
Using this, it is easy to show that
\begin{equation}
    \left[f_\rho(\bm{x}) + f_{-\rho}(\bm{x})\right]^\text{DB} =
    \frac{\omega_\rho(\bm{x}) - \omega_{-\rho}(\bm{x})}{\log \left(\omega_\rho(\bm{x})/\omega_{-\rho}(\bm{x})\right)}.
    \label{eq:symmetric_f}
\end{equation}
This result implies that employing the diffusion tensor
\begin{equation}
    D_{nm}^\text{DB}(\bm{x}) \equiv \frac{1}{\Omega^2}\sum_{\rho > 0} \frac{\lambda_\rho(\bm{x})-\lambda_{-\rho}(\bm{x})}{\log \left(\lambda_\rho(\bm{x})/\lambda_{-\rho}(\bm{x})\right)}\Delta_\rho^n \Delta_\rho^m
    \label{eq:equilibrium_diff}
\end{equation}
in Eq.~\eqref{eq:FP} corresponds to a diffusive process that is guaranteed to have the same macroscopic fluctuations as the original Markov jump process.

The previous result can also be stated in connection to the Einstein-Smoluchowski relation for general diffusive dynamics. This relation just expresses the fact that for a Fokker-Planck dynamics like the one in Eq.~\eqref{eq:FP} to relax to the Gibbs distribution corresponding to a potential $\Phi(\bm{x})$, the drift vector $\mu_n(\bm{x})$ and diffusion tensor $D_{nm}(\bm{x})$ must be related to each other by the identity $\mu_n(\bm{x}) = -D_{nm}(\bm{x}) \partial_m \Phi(\bm{x}) + \partial_m D_{mn}(\bm{x})$. Indeed, from Eq.~\eqref{eq:equilibrium_diff} it follows that:
\begin{align}
    &\lim_{\Omega\to\infty} \bigg(\mu_n(\bm{x}) + D^\text{DB}_{nm}(\bm{x}) \partial_m \Phi(\bm{x}) - \partial_m D^\text{DB}_{mn}(\bm{x})\bigg)\nonumber\\
    &= \sum_{\rho>0} \left[  (\omega_\rho - \omega_{-\rho})  + \frac{\omega_\rho-\omega_{-\rho}}{\log( \omega_\rho/\omega_{-\rho})}(\bm{\Delta}_\rho\cdot \nabla \phi) \right] \Delta_\rho^n  = 0\label{eq:einstein-rel}
\end{align}
where in the last line we used the macroscopic limit of the detailed balance conditions, Eq.~\eqref{eq:macro-db}. We note that the usual Kramers-Moyal diffusion tensor in Eq.~\eqref{eq:kramers_moyal_diff_3} fails to satisfy the previous condition. In particular, it is well documented that the Kramers--Moyal diffusion tensor can yield spurious steady currents and hence a Fokker--Planck dynamics that \emph{violates} detailed balance even when the underlying MJP is at equilibrium \cite{mendler2020, ceccato2018}. By contrast, Eq. \eqref{eq:einstein-rel} demonstrates that our construction maps equilibrium MJPs to equilibrium diffusions, a property that is highly desirable for any diffusion approximation.
\\

We now show in what follows that the diffusion tensor in Eq.~\eqref{eq:equilibrium_diff} also leads to the correct macroscopic fluctuations for first-order deviations from perfect detailed-balance conditions. This is a crucial result of our theory since it significantly extends its scope of validity.

\begin{figure}
    \centering
    \includegraphics[width=0.99\linewidth]{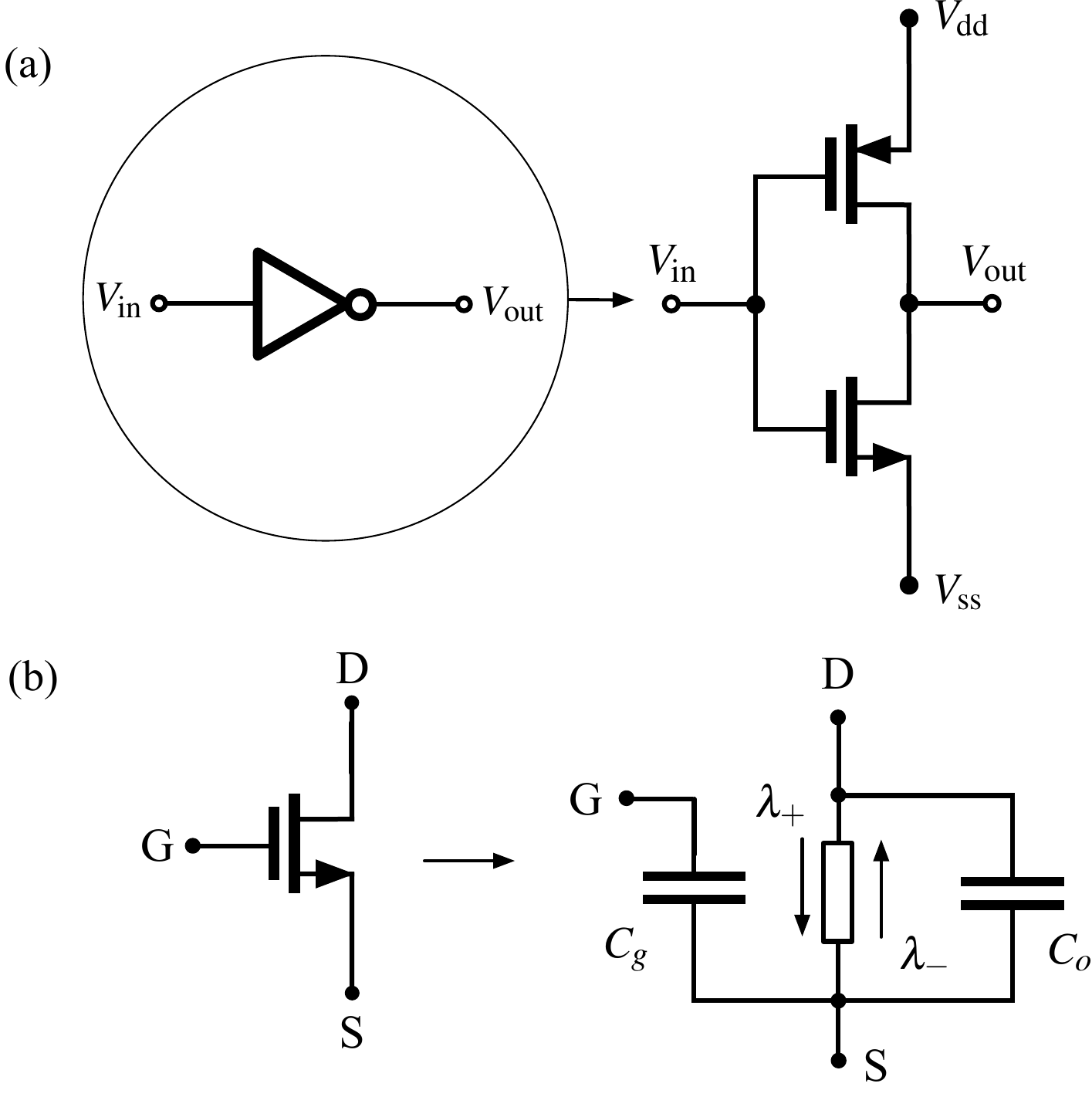}
    \caption{{\bf Stochastic CMOS circuits}. (a) In the context of digital electronics, a logical NOT gate (left) is typically implemented as a CMOS inverter (right), which employs two field-effect transistors. In the examples that follow, $V_\text{in} = 0$ and $V_\text{dd}= -V_\text{ss} = V_T$, with $V_T \equiv k_BT/e$ the thermal voltage. (b) Each transistor is modeled in the subthreshold regime as a stochastic conduction channel between its source (S) and drain (D) terminals, with voltage-dependent Poisson rates $\lambda_{+/-}$, cf. \cite{freitasStochastic2021} for a detailed discussion. For the circuit examples used in this paper, $C_o=2C_T$ and $C_g = 4C_T$, where $C_T \equiv e/V_T$. }
    \label{fig:CMOS-explainer}
\end{figure}

\section{Dynamical large-deviations theory and linear response regime}
\label{sec:dynamical-ldt}

In this section we consider the linear response regime around detailed-balance conditions and sketch the main ingredients of the proof showing that the previous results are still valid in that regime. We first collect relevant results concerning the equations of motion satisfied by the large-deviation rate functions $I(\bm{x},t)$, introduced in Eq.~\eqref{eq:self-information-expansion}. For Markov jump processes, the rate function evolves according to the equation of motion \cite{kubo1973fluctuation}
\begin{align}
    \partial_t I(\bm{x},t) = \sum_\rho \omega_\rho(\bm{x}) \left(1-e^{\bm{\Delta}_\rho\cdot \nabla I(\bm{x},t)} \right).
    \label{eq:MEQ-rate-function}
\end{align}
On the other hand, the rate function for a drift-diffusion process evolves according to the equation of motion \cite{santolin2024}
\begin{align}
    \partial_t I(\bm{x},t)& = u_n(\bm{x}) \partial_n I(\bm{x},t)\nonumber\\
    &~~~~~~~~~~~~~~~+ d_{nm}(\bm{x}) \partial_n I(\bm{x},t) \partial_m I(\bm{x},t),
    \label{eq:FP-rate-function}
\end{align}
which is written in terms of the following macroscopic versions of the drift vector and diffusion tensor \cite{santolin2024}: 
\begin{align}
    u_n(\bm{x})&\equiv \lim_{\Omega \to\infty}\mu_n(\bm{x})\\
    d_{nm}(\bm{x})&\equiv \lim_{\Omega\to\infty} \Omega D_{nm}(\bm{x})
    \label{eq:macro_drift_diff}
\end{align}
To dominant order in $\Omega$ and for detailed-balance conditions (Eq.~\eqref{eq:macro-db}), the Einstein-Smoluchowski relation in Eq.~\eqref{eq:einstein-rel} reads as the following relation between the above macroscopic versions of the drift vector and diffusion tensor:
\begin{align}
    u(\bm{x}) = -d^\text{EQ}(\bm{x})\cdot \nabla \phi(\bm{x}).\label{eq:einstein-rel2}
\end{align}
In particular, we can substitute the above into Eq.~\eqref{eq:FP-rate-function} and confirm that $I_\text{eq}(\bm{x}) = \phi(\bm{x})$ is a steady-state solution of both equations of motion. Therefore, confirming the previous results, our new diffusion approximation yields the correct macroscopic steady-state potential at equilibrium.

\subsection{Linear response}

In the linear response regime, we consider small departures from the global detailed balance condition. In particular, the detailed balance conditions for the underlying MJP in  Eq.~\eqref{eq:macro-db} are generalized to the {\it local} detailed balance condition
\begin{align}
    \log \frac{\omega_\rho(\bm{x})}{\omega_{-\rho}(\bm{x})} = -\bm{\Delta}_\rho\cdot \nabla \phi(\bm{x}) + \epsilon w_\rho(\bm{x}),
\end{align}
for $w_\rho(\bm{x})$ some function of both the state and the transition, and $\epsilon$ a perturbative parameter that allows us to continuously tune to the detailed-balanced limit $\epsilon\to 0$. In Sec. I of the SM, we show that, under mild conditions on $u(\bm{x})$, the steady-state rate function $I_\text{ss}(\bm{x})$ for the diffusive approximation of Eq.~\eqref{eq:improved_diff} agrees with the corresponding rate function of the underlying Markov jump process to first-order in $\epsilon$. The proof exploits recent results on macroscopic stochastic thermodynamics \cite{falasco2023} and can be intuitively understood as follows. As shown in \cite{freitas2021linear, freitas2022emergent} for MJPs and in \cite{santolin2024} for diffusive processes, the steady-state rate function can be directly related to the macroscopic entropy production rate in the linear response regime. In both cases, this relation reads 
\begin{equation}
d_t I_\text{ss}(\bm{x}_t) = -\dot \sigma(\bm{x}_t) + \mathcal{O}(\epsilon^2),
\end{equation}
where $\dot\sigma(\bm{x})$ is the scaled macroscopic entropy production rate and $\bm{x}_t$ is any trajectory satisfying the deterministic equation of motion $d_t \bm{x}_t = u(\bm{x}_t)$. The difference between MJPs and diffusive processes lies in their respective expressions for the entropy production rate $\dot \sigma(\bm{x})$. For MJPs we have 
\begin{equation}
    \dot \sigma_\text{MJP}(\bm{x}) = \sum_{\rho>0} (\omega_\rho(\bm{x}) - \omega_{-\rho}(\bm{x}))
    \log\frac{\omega_\rho(\bm{x})}{\omega_{-\rho}(\bm{x})},
    \label{eq:epr_MJP}
\end{equation}
while for diffusive processes
\begin{equation}
    \dot \sigma_\text{diff}(\bm{x}) = u_n(\bm{x}) d_{nm}(\bm{x}) u_m(\bm{x}).
\end{equation}
It is possible to show that if $d_{nm}(\bm{x})$ follows from Eq.~\eqref{eq:improved_diff}, then $\int dt \: \sigma_\text{MJP}(\bm{x}_t) = \int dt \: \sigma_\text{diff}(\bm{x}_t) + \mathcal{O}(\epsilon^2)$ for any deterministic trajectory $\bm{x}_t$, which then implies that 
$I^\text{MJP}_\text{ss}(\bm{x}) = I^\text{diff}_\text{ss}(\bm{x}) + \mathcal{O}(\epsilon^2)$. See the SM, Sec. I for an explicit perturbative proof. Therefore, we conclude that $I^\text{MJP}_\text{ss}(\bm{x})$ and $I^\text{diff}_\text{ss}(\bm{x})$ match each other to first order in $\epsilon$. Of course, this is not the case for a diffusive process based on the usual Kramers-Moyal approximation of Eq.~\eqref{eq:kramers_moyal_diff_1}, since in that case both rate functions differ even for $\epsilon=0$.

\section{Examples}
\label{sec:examples}

We now focus on example systems with which to compare our new diffusive approximation with the standard Kramers-Moyal expansion. We will see in each example that the improved diffusive approximation more accurately captures phenomena associated with rare-events/large steady state fluctuations. Throughout this section we will focus on examples of Markov jump processes coming from electronic circuits that commonly appear in complementary metal-oxide semiconductor (CMOS) technology, but our formalism is more general and applies to generic Markov jump processes, e.g. chemical reaction networks, random walks, etc.\\

\begin{figure}
    \centering
    \includegraphics[width=0.99\linewidth]{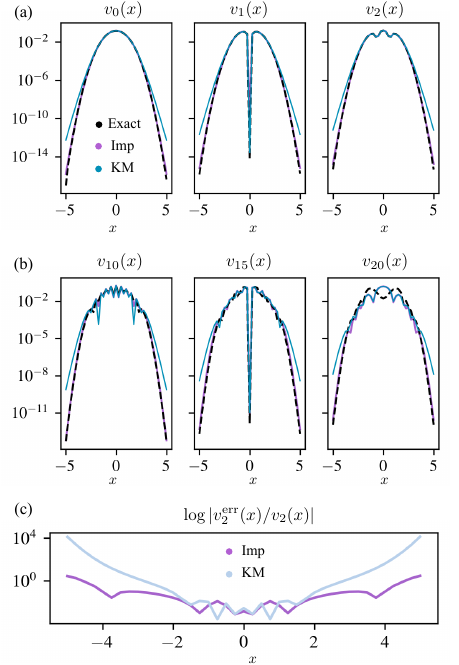}
    \caption{{\bf Liouvillian eigenmodes of a CMOS inverter under different diffusive approximations}. (a) Three slowest-decaying eigenvectors for the Liouvillian of the inverter circuit in Figure \ref{fig:CMOS-explainer}a, plotted together with the three slowest-decaying eigenmodes of the Liouvillian for both the Kramers-Moyal approximation and the improved diffusion approximation. (b) Representative mid--spectrum eigenmodes (11th, 16th, and 21st slowest) for the same system.
 (c) Fractional eigenmode error $v_2^\text{err}(x) \equiv v_2(x)-v_2^\text{exact}(x)$, for both the Kramers-Moyal and improved diffusive approximations. Parameters are the same as in panel (a).}
    \label{fig:CMOS-inverter}
\end{figure}

\subsection{One-dimensional jump processes}

Consider a one-dimensional Markov jump process with nearest-neighbor jumps $\Delta_\pm = \pm 1$. In this simple case, we can analyze the performance of both diffusive approximations analytically. Indeed, with only two distinct jump vectors, the sum in Eq.~\eqref{eq:MEQ-rate-function} has only two terms, and we immediately obtain the solution $I(x) = -\int dx\,\log\frac{\omega_+}{\omega_-}$. On the other hand, for a one-dimensional drift-diffusion process, Eq.~\eqref{eq:FP-rate-function} admits the steady-state solution $I(x) = -\int dx\, \frac{u}{d}$. The diffusion tensor for the improved diffusive approximation satisfies $d = -u\log (\omega_+/\omega_-)$ and thus \cite{hanggi1988bistability}
\begin{align}
    \frac{d}{dx}I_\text{IMP}(x)  = -\log \frac{\omega_+(x)}{\omega_-(x)},
\end{align}
which implies that, in this simple case, the steady-state large-deviation function obtained by the improved diffusion approximation is exact. The reason is that this kind of 1D system can always be considered to be detailed-balance. In contrast, repeating the same exercise for the Kramers-Moyal diffusion approximation yields 
\begin{align}
    \frac{d}{dx}I_\text{KM}(x) = -2\frac{\omega_+(x) - \omega_-(x)}{\omega_+(x) +\omega_-(x)}.
\end{align}
These results are analogous to those obtained in \cite{hanggi1984, hanggi1988bistability}.
To quantify precisely by how much the Kramers-Moyal approximation breaks down, we expand both rate functions near a deterministic fixed point $x^{*}$, for which $\omega_+(x^*)=\omega_-(x^*)$:
\begin{align}
    \frac{dI_\text{IMP}}{dx}-\frac{dI_\text{KM}}{dx} =\sum_{k>0}\bigg(\frac{1}{k} - \frac{1}{2^{k-1}}\bigg)\bigg(1-\frac{\omega_+}{\omega_-}\bigg)^k.
\end{align}
Because the above difference is of cubic order in $1-\omega_+/\omega_-$, to see a difference between the two diffusive approximations we must evaluate the steady-state distribution far away from any deterministic fixed point. This means that we must look at the tails of the distribution to see a difference between the approximations. We will see this behavior in the example discussed below.

\begin{figure*}
    \centering
    \includegraphics[width=0.99\linewidth]{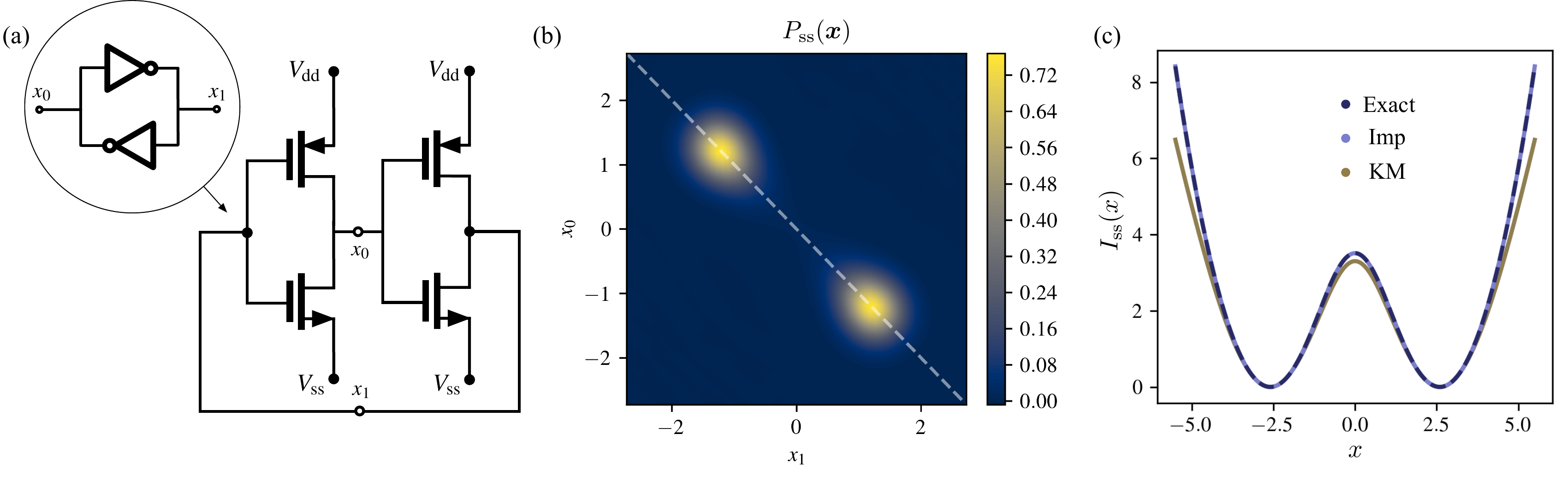}
    \caption{(a) Circuit diagram for a random-access memory (RAM) cell formed by two CMOS inverters connected in a feedback loop (left inset). The stochastic degrees of freedom here are the voltages $x_0,x_1$ at the two output terminals of the circuit. We consider $V_\text{dd} = -V_\text{ss}$ for the powering voltages. (b) Steady-state probability density for the circuit in panel (a), under the powering configuration $V_\text{dd} =-V_\text{ss} = 1.3V_T$, and with $\Omega = 11$. The bistable axis $x_1\equiv -x_0$ for the memory is marked with a dashed line. (c) Cut along the bistable axis of the large-deviation function $I_\text{ss}(\bm{x})$ for the circuit. Here, $V_\text{dd} = -V_\text{ss} =2.6V_T$. We show results obtained from both the Kramers-Moyal approximation as well as the improved diffusion approximation. }
    \label{fig:CMOS-pbit}
\end{figure*}

\subsubsection{CMOS inverter}
As a basic example of a one-dimensional jump process we consider a CMOS inverter, cf. Figure \ref{fig:CMOS-explainer}a. In the inverter circuit, an $n$- and $p$-type field-effect transistor are connected in series and a common input voltage $V_\text{in}$ is applied to the gate of each transistor (in this case $V_\text{in}=0$). When both transistors are operated in the subthreshold regime, the dimensionless voltage $x\equiv V_\text{out}/V_T$ at the output node (in units of the thermal voltage $V_T\equiv k_BT/e$) can be modelled as a Markov jump process with nearest-neighbor jumps $\Delta_\pm = \pm 1$ and jump rates
\begin{align}
    \lambda_+(x)&= \frac{1}{\tau}\bigg(e^{-(x+\Delta x)}e^{-(1 + 1/n)V_\text{dd}/V_T} + e^{+V_\text{dd}/nV_T}\bigg)\nonumber\\
    \lambda_-(x)&=\frac{1}{\tau}\bigg(e^{+(x - \Delta x)}e^{-(1 - 1/n)V_\text{dd}/V_T} + e^{-V_\text{dd}/nV_T}\bigg)
\end{align}
Here $\Delta x\equiv (2\Omega)^{-1}$ is half the change in the dimensionless voltage $x$ after a single jump, and $n$ is a constant called the subthreshold slope. Furthermore, $V_\text{dd}$ is the powering voltage of the logic gate (cf. Figure \ref{fig:CMOS-explainer}a), and $\tau$ a time constant. In this context of stochastic circuits, $\Omega$ is the smallest eigenvalue of the dimensionless matrix $C/C_T$, where $C$ is the Maxwell capacitance matrix for the circuit, with $C_T \equiv e^2/k_BT$ an elementary unit of capacitance.\\

Issues with the Kramers-Moyal approximation in accurately capturing large stationary voltage fluctuations in this circuit were noted in \cite{gopal2022}. However, this observation was limited to the steady state and did not cover quantities relevant to transient dynamics, such as higher-order eigenmodes. In our numerical analysis of this example (as well as all following examples) we analyze the full spectral properties of the Liouvillian $\mathcal L$ for our Markov jump process and its diffusive approximations. In terms of the spectral decomposition of $\mathcal L$, the evolution of an initial distribution $P(\bm{x},0)= \sum_k c_k v_k(\bm{x})$ takes the form
\begin{align}
    P(\bm{x},t) = \sum_k c_k e^{-\lambda_kt} v_k(\bm{x}),
\end{align}
where $\lambda_k$ and $v_k(\bm{x})$ are the eigenvalues and right eigenmodes of $\mathcal{L}$. To solve this problem numerically, we truncate the state space to a finite box $x\in [-L, L ]^{\times d}$, where $L$ is some large cutoff voltage, and impose reflecting boundary conditions on $\mathcal L$. For diffusive dynamics we also discretize the state space using a sufficiently fine grid (see the SM, Sec. IV for more details). We then use a sparse solver to obtain the low-lying eigenmodes and eigenvalues $v_k, \lambda_k$ for the resulting finite-dimensional problem. We then compare the results with the eigenvectors and eigenvalues $\lambda_k^\text{exact},v_k^\text{exact}$ of the Liouvillian for the original Markov jump process.\\

In Figure \ref{fig:CMOS-inverter}a we plot the three slowest decaying eigenmodes $v_0,v_1,v_2$ of the Liouvillian for the inverter circuit. In confirmation of our analytic arguments, we observe that the improved diffusive approximation more accurately captures the tails of the steady state distribution far away from the deterministic fixed point $x^*=0$. However, from the numerics we also observe improvements in the tails of fast-decaying eigenmodes above the steady state, e.g. $v_{10}, v_{15}, v_{20}$, cf. Figure \ref{fig:CMOS-inverter}b, a result that goes beyond the predictions of our analytic arguments.\\

\subsection{Probabilistic bit (CMOS implementation)}
We now consider an example of a two-dimensional system that, at variance with the 1D examples above and in \cite{hanggi1984, hanggi1988bistability}, is genuinely non-detailed-balanced. It consists of a bistable circuit formed by two CMOS inverters connected in feedback, cf. Figure \ref{fig:CMOS-pbit}a. Again, we use the model developed in \cite{freitasStochastic2021}, under which the normalized voltages $\bm{x} \equiv (V_0/V_T, V_1/V_T)$ at the output terminals of the circuit evolve according to a Markov jump process with jump vectors $\Delta_{j,\pm}^n = \pm \delta_{j,n}$, with corresponding jump rates \cite{freitasReliability2022}
\begin{align}
    \lambda_{0,\pm}(\bm{x})&=\frac{1}{\tau}\bigg(e^{V_\text{dd}/nV_T\mp x_1/n} + e^{\pm x_1/n}e^{\mp (x_0\pm \Delta x)}\bigg)\\
    \lambda_{1,\pm}(\bm{x})&=\frac{1}{\tau}\bigg(e^{V_\text{dd}/nV_T\mp x_0/n} + e^{\pm x_0/n}e^{\mp (x_1\pm \Delta x)}\bigg)
\end{align}
Here, $V_\text{dd}$ is the powering voltage of the inverters in the memory cell (cf. Figure \ref{fig:CMOS-pbit}), and as before $\tau$ is a time constant. 

This model has a transition from a monostable to a bistable steady-state at the critical powering bias $V_\text{dd} = \log(2)V_T$ \cite{freitasReliability2022}.  In Figure \ref{fig:CMOS-memory}a we plot, as a function of the powering voltage and for different scale parameters $\Omega$, the Kullback-Leibler (KL) divergence of the steady-state distribution of each diffusive approximation with respect to the exact steady-state of the MJP. We see that the improved diffusion approximation results in a lower KL divergence for all powering voltages and scale parameters. Notably, in the bistable regime $V_\text{dd} > \log(2)V_T$, the KL divergence increases with the powering voltage for the Kramers-Moyal approximation while it levels off for the improved diffusion approximation. Thus, the improved diffusion approximation leads to a better description of steady-state fluctuations even for non-detailed-balance conditions.

We now turn our attention to the behaviour of the smallest non-zero eigenvalue $\lambda_1$ of the Liouvillian of both diffusive approximations. In the bistable phase, this eigenvalue gives the rate of thermally induced switching transitions between the two metastable states, i.e., the rate of logical errors in the memory. In Figure \ref{fig:CMOS-memory}b-c we show the relative difference of $\lambda_1$ with respect to the corresponding eigenvalue of the MJP. While the Kramers-Moyal approximation shows a better performance in the monostable regime $V_\text{dd} < \log(2)V_T$, the improved diffusion approximation shows a much slower increase of the error in the bistable regime $ V_\text{dd} > \log(2)V_T$. Furthermore, as we show in more detail in Figure \ref{fig:CMOS-error-rate}, the performance of the improved diffusion approximation becomes insensitive to the scale. In contrast, the performance of the Kramers-Moyal approximation degrades 
as one goes deeper in the macroscopic limit. As we show below, this can be understood in terms of a generalization of the Eyring--Kramers formula to NESSs \cite{bouchetGeneralization2016}.

\subsubsection{Reduced rate-function and metastable lifetime of the memory}

The bistable voltage states for this circuit lie along the line $x_1 = -x_0$ \cite{freitasStochastic2021}. The rate function for this circuit for CMOS memory can be restricted to this line, and admits the solution \cite{freitasReliability2022}
\begin{align}
    I(x) = -2\int dx\,\log \frac{\omega(x)}{\omega(-x)}
\end{align}
where $I(x)\equiv I(x,-x)$ and $\omega(x) \equiv \omega_{0,+}(x,-x)$. In deriving this identity, one uses the symmetries that relate the jump rates for this circuit, as well as the fact that the restricted rate function $f(y) \equiv I(y+x, y-x)$ always attains a global minimum at $y = 0$.\\

\begin{figure}
    \centering
    \includegraphics[width=1.00\linewidth]{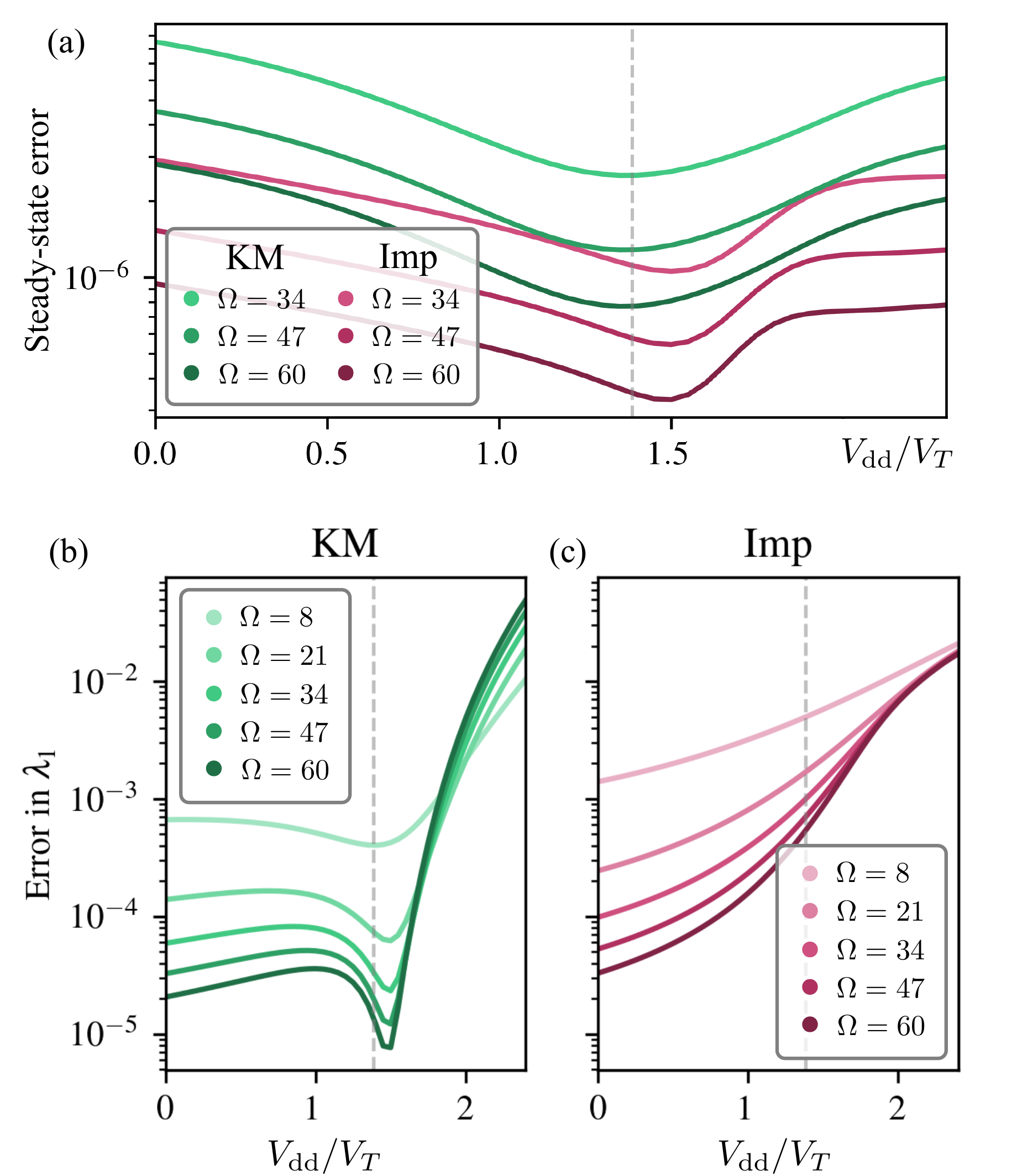}
    \caption{{\bf Performance of improved diffusion approximation applied to a CMOS memory}. (a) Error (as measured by the Kullback-Leibler divergence) in estimating the nonequilibrium steady-state of a CMOS random-access memory (RAM) cell (cf. Figure \ref{fig:CMOS-pbit}a), under the Kramers-Moyal and improved diffusion approximations, as a function of the powering voltage $V_\text{dd}$ for the circuit, in units of $V_T \equiv k_BT/e$. Here, $\Omega$ corresponds to the smallest eigenvalue of the Maxwell capacitance matrix of the underlying circuit. (b) Fractional error in $\lambda_1$, the first Liouvillian eigenvalue, for both approximations. At $V_\text{dd} = \log (2)\, V_T$ (vertical dashed line), the system undergoes a phase transition and becomes bistable. }
    \label{fig:CMOS-memory}
\end{figure}

We can apply the same assumptions to the rate function of any diffusive approximation for the dynamics of the circuit, in which case we find that Eq.~\eqref{eq:FP-rate-function} simplifies to
\begin{align}
   u_\text{eff}\frac{dI}{dx}  + d_\text{eff}\bigg(\frac{dI}{dx}\bigg)^2 = 0
\end{align}
where $u_\text{eff} = (u_0 - u_1)/2$ and $d_\text{eff} 
 = (d_{00} + d_{11})/4$ are effective drift and diffusion coefficients along the bistable axis $y\equiv 0$ of the memory. Substituting in the objects from the improved diffusion approximation, we obtain the solution
 \begin{align}
     \frac{dI_\text{IMP}}{dx} = -2\log \frac{\omega(x)}{\omega(-x)},
\end{align}
 which means that $I_\text{IMP}$ and $I$ exactly coincide along the bistable line $x_1 = -x_0$. Repeating this calculation for the Kramers-Moyal approximation, we obtain $\frac{dI_\text{KM}}{dx} = -4\frac{\omega(x) - \omega(-x)}{\omega(x) + \omega(-x)}$. In particular, the rate function given by the Kramers-Moyal diffusion approximation is exact up to second-order in the expansion parameter $1-\omega(x)/\omega(-x)$ (as was seen in the example of the CMOS inverter). However, at high powering voltages, i.e. $V_{dd} \gtrsim 2V_T$, the discrepancy is enough to produce an observable difference in the value of the rate function at the center point $x=0$, cf. Figure \ref{fig:CMOS-pbit}c.\\

The first Liouvillian eigenvalue for our circuit can be estimated from the stationary rate function $I$ via the following expression \cite{bouchetGeneralization2016},
\begin{align}
    \lim_{\Omega\to \infty} \Omega^{-1}\log \lambda_1 = \Delta I,
\end{align}
where $\Delta I$ is the difference of the stationary rate function between the center point $x=0$ and any of its two minima. This result is the non-equilibrium generalization of the well-known Eyring--Kramers formula for metastable transitions in detailed-balanced systems \cite{bouchetGeneralization2016}.
Because $ I_\text{KM}\neq I$ whereas $ I_\text{IMP}=  I$, we expect the improved diffusion approximation to provide a better estimate for $\lambda_1$ in the large system size limit, whenever $\Delta I_\text{KM}$ is appreciably different from $\Delta I$. $\Delta I_\text{KM}$ is appreciably different from $\Delta I$ only at powering voltages $V_\text{dd} \gtrsim 2V_T$, and accordingly we see the improved approximation outperform the Kramers-Moyal approximation when the memory is operated at these higher powering voltages, cf. Figure \ref{fig:CMOS-memory}b. \\

\begin{figure}
    \centering
    \includegraphics[width=1.00\linewidth]{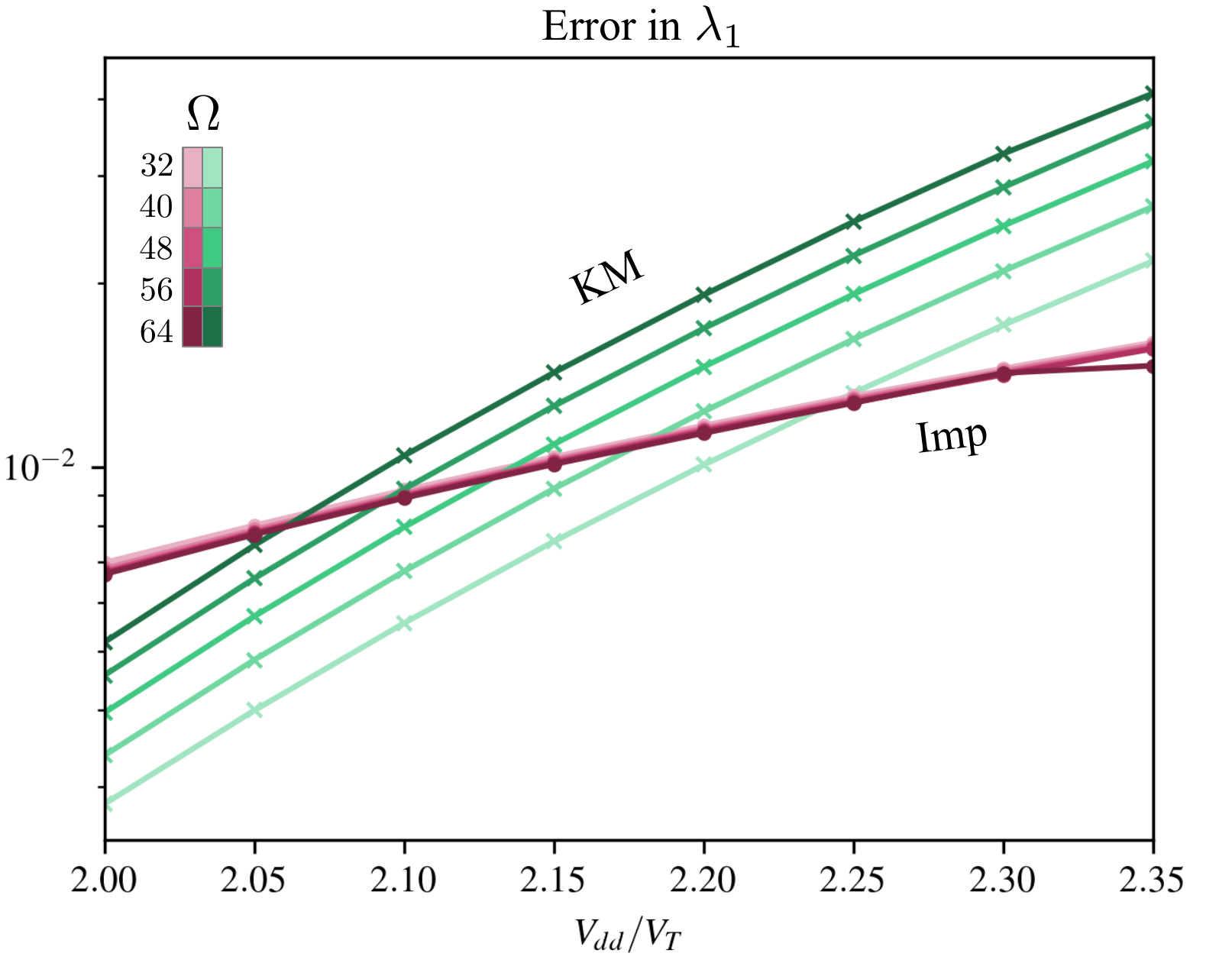}
    \caption{{\bf Error in estimating the error-rate of a CMOS memory cell}. Same data as in Figure \ref{fig:CMOS-memory}b, except with a slightly-higher range for $\Omega$ and $V_\text{dd}$.}
    \label{fig:CMOS-error-rate}
\end{figure}

\begin{figure}
    \centering
    \includegraphics[width=0.99\linewidth]{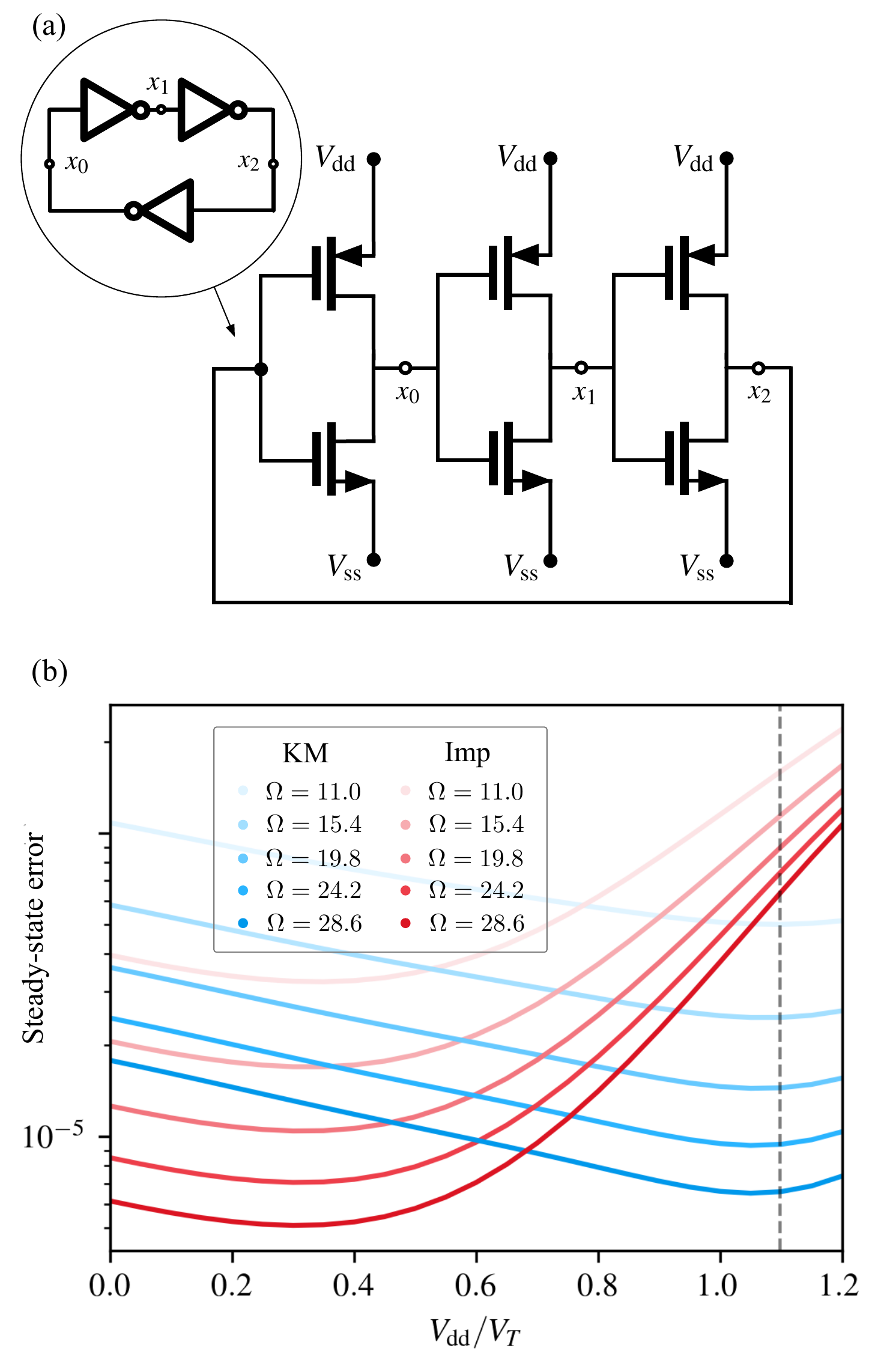}
    \caption{ {\bf Error in estimating the nonequilibrium steady-state of a CMOS ring oscillator.} (a) A ring oscillator consists of an odd number of inverters connected in feedback. Here, we show a CMOS implementation of a ring oscillator with three inverters. (b) Error (as measured by the Kullback-Leibler divergence) in estimating the nonequilibrium steady-state of the ring oscillator as a function of the powering voltage of the oscillator (darker curves indicate increasing $\Omega$). Here, $V_\text{ss} = -V_\text{dd}$. Above the threshold voltage $V_\text{dd}/V_T = \log\,3$ (dashed vertical line), the system undergoes a phase transition and a limit cycle emerges in the deterministic dynamics for the oscillator. }
    \label{fig:CMOS-ring-oscillator}
\end{figure}

\subsection{Fixed-points vs limit cycles}

Although the improved diffusion approximation is theoretically expected to work well only in the linear-response regime close to detailed-balance conditions, the previous example shows that it can  outperform the usual Kramers-Moyal approximation even beyond linear order in the thermodynamic forces that break detailed balance. This can be understood as follows. The deterministic dynamics of the CMOS memory has fixed-point attractors $\bm{x}^*$ at which the drift vanishes: $\mu_n(\bm{x}^*) = 0$. For cases in which the different jump vectors $\bm{\Delta}_\rho$ are linearly independent, which is a condition satisfied in the examples considered, $\mu_n(\bm{x}^*) = 0$ implies that $\lambda_\rho(\bm{x}^*) - \lambda_{-\rho}(\bm{x}^*) = 0$ for all $\rho$. This in turn implies that the entropy production of the original MJP vanishes at the fixed points (see Eq.~\eqref{eq:epr_MJP}). Thus, near the fixed points, where most of the probability is concentrated in the macroscopic limit, the system operates close to detailed-balanced conditions, even if detailed-balance is significantly broken at a global level. For this reason, we expect the better performance of the improved diffusion approximation to carry on to other systems satisfying the above conditions.

However, for systems that have more complex dynamical attractors at which entropy is produced, we do not expect the advantages of the improved diffusion approximation to extend beyond the linear-response regime. To test this case, we have performed simulations on the CMOS ring oscillator shown in Figure \ref{fig:CMOS-ring-oscillator}a. This circuit can be described via a MJP in the same way as the examples considered above. It features a Hopf bifurcation from a monostable phase into a limit cycle at $V_\text{dd} = -V_\text{ss} = \log(3) V_T$. As with the CMOS memory, we numerically compute the steady-state distributions according to both diffusive approximations, and measure their KL divergence with respect to the exact steady-state of the MJP. The results are shown in Figure \ref{fig:CMOS-ring-oscillator}b. We indeed see that, as expected from the theory, the improved diffusion approximation only performs better than the usual Kramers-Moyal approximation for moderate powering voltages, below the transition into a limit cycle.

\section{Discussion}
\label{sec:discussion}

In this work, we have considered an alternative diffusive approximation of Markov jump processes that, under strict detailed-balance conditions, was known to outperform the usual diffusive approximation based on the truncation to second order of the Kramers-Moyal expansion. We have extended the validity of this method by proving that its main advantage, namely the accurate description of arbitrarily large macroscopic steady-state fluctuations, is preserved in the linear-response regime around strict detailed-balance. By comparing the performance of both diffusive approximations on stochastic models of realistic electronic circuits, we have shown that the improved diffusion approximation is also more accurate for the description of dynamical features beyond steady-state fluctuations, in particular higher order dynamical modes and relaxation times. The theoretical arguments and numerical experiments supporting our results indicate that these advantages can be expected to hold significantly beyond the linear-response regime for systems with fixed-point attractors, while not for systems with more complex dynamical attractors like limit cycles.


\bibliography{references.bib}

@article{freitas2025,
    title     = {\emph{Taming non-equilibrium thermal fluctuations in subthreshold CMOS circuits}},
    author    = {Nahuel Freitas and Geremia Massarelli and Jeremy Rothschild and Dylan Keane and Ethan Dawe and Sewook Hwang and Akhil Garlapati and Trevor McCourt},
    journal   = {Phys. Rev. Lett.},
    year      = {2025},
    note      = {{Submitted}}
}

@article{jelincic2025,
    title     = {\emph{An efficient probabilistic hardware architecture for diffusion-like models}},
    author    = {Andraž Jelinčič and Owen Lockwood and Sewook Hwang and Akhil Garlapati and Guillaume Verdon and Trevor McCourt},
    journal   = {Phys. Rev. X},
    year      = {2025},
    note      = {{Submitted}}
}

@article{challis2016,
  title = {Numerical study of the tight-binding approach to overdamped Brownian motion on a tilted periodic potential},
  author = {Challis, K. J.},
  journal = {Phys. Rev. E},
  volume = {94},
  issue = {6},
  pages = {062123},
  numpages = {10},
  year = {2016},
  month = {Dec},
  publisher = {American Physical Society},
  doi = {10.1103/PhysRevE.94.062123},
  url = {https://link.aps.org/doi/10.1103/PhysRevE.94.062123}
}

@article{hayashi2015,
  title = {Giant Acceleration of Diffusion Observed in a Single-Molecule Experiment on ${\mathrm{F}}_{1}\text{\ensuremath{-}}\mathrm{ATPase}$},
  author = {Hayashi, Ryunosuke and Sasaki, Kazuo and Nakamura, Shuichi and Kudo, Seishi and Inoue, Yuichi and Noji, Hiroyuki and Hayashi, Kumiko},
  journal = {Phys. Rev. Lett.},
  volume = {114},
  issue = {24},
  pages = {248101},
  numpages = {5},
  year = {2015},
  month = {Jun},
  publisher = {American Physical Society},
  doi = {10.1103/PhysRevLett.114.248101},
  url = {https://link.aps.org/doi/10.1103/PhysRevLett.114.248101}
}

@article{ceccato2018,
    author = {Ceccato, Alessandro and Frezzato, Diego},
    title = {Remarks on the chemical Fokker-Planck and Langevin equations: Nonphysical currents at equilibrium},
    journal = {The Journal of Chemical Physics},
    volume = {148},
    number = {6},
    pages = {064114},
    year = {2018},
    month = {02},
    issn = {0021-9606},
    doi = {10.1063/1.5016158},
    url = {https://doi.org/10.1063/1.5016158},
}

@article{mendler2020,
  title = {Predicting properties of the stationary probability currents for two-species reaction systems without solving the Fokker-Planck equation},
  author = {Mendler, Marc and Drossel, Barbara},
  journal = {Phys. Rev. E},
  volume = {102},
  issue = {2},
  pages = {022208},
  numpages = {13},
  year = {2020},
  month = {Aug},
  publisher = {American Physical Society},
  doi = {10.1103/PhysRevE.102.022208},
  url = {https://link.aps.org/doi/10.1103/PhysRevE.102.022208}
}

@article{holek2020,
    author = {Santamaría-Holek, I. and Pérez-Madrid, A.},
    title = {Eyring equation and fluctuation–dissipation far away from equilibrium},
    journal = {The Journal of Chemical Physics},
    volume = {153},
    number = {24},
    pages = {244116},
    year = {2020},
    month = {12},
    issn = {0021-9606},
    doi = {10.1063/5.0032634},
    url = {https://doi.org/10.1063/5.0032634},
}

@article{panahi2025,
author = {Panahi, Shirin  and Feudel, Ulrike  and Abbott, Karen C.  and Hastings, Alan  and Lai, Ying-Cheng },
title = {Generalized paradox of enrichment: noise-driven rare rarity in degraded ecological systems},
journal = {Journal of The Royal Society Interface},
volume = {22},
number = {230},
pages = {20250087},
year = {2025},
doi = {10.1098/rsif.2025.0087},

URL = {https://royalsocietypublishing.org/doi/abs/10.1098/rsif.2025.0087}
}

@article{alamilla2020,
  title = {Enhanced diffusion and the eigenvalue band structure of Brownian motion in tilted periodic potentials},
  author = {L\'opez-Alamilla, N. J. and Jack, M. W. and Challis, K. J.},
  journal = {Phys. Rev. E},
  volume = {102},
  issue = {4},
  pages = {042405},
  numpages = {9},
  year = {2020},
  month = {Oct},
  publisher = {American Physical Society},
  doi = {10.1103/PhysRevE.102.042405},
  url = {https://link.aps.org/doi/10.1103/PhysRevE.102.042405}
}

@article{gopal2022,
  title = {\emph{Large deviations theory for noisy nonlinear electronics: CMOS inverter as a case study}},
  author = {Gopal, Ashwin and Esposito, Massimiliano and Freitas, Nahuel},
  journal = {Phys. Rev. B},
  volume = {106},
  issue = {15},
  pages = {155303},
  numpages = {20},
  year = {2022},
  month = {Oct},
  publisher = {American Physical Society},
  doi = {10.1103/PhysRevB.106.155303},
  url = {https://doi.org/10.1103/PhysRevB.106.155303}
}

@article{Lebowitz1999,
	author = {Lebowitz, Joel L. and Spohn, Herbert},
	title = {\emph{A Gallavotti{\textendash}Cohen-Type Symmetry in the Large Deviation Functional for Stochastic Dynamics}},
	journal = {J. Stat. Phys.},
	volume = {95},
	number = {1},
	pages = {333--365},
	year = {1999},
	month = apr,
	issn = {1572-9613},
	publisher = {Kluwer Academic Publishers-Plenum Publishers},
	doi = {10.1023/A:1004589714161},
    url = {https://doi.org/10.1023/A:1004589714161}
}

@article{Spohn1983,
	author = {Spohn, H.},
	title = {\emph{Long range correlations for stochastic lattice gases in a non-equilibrium steady state}},
	journal = {J. Phys. A: Math. Gen.},
	volume = {16},
	number = {18},
	pages = {4275},
	year = {1983},
	month = dec,
	issn = {0305-4470},
	publisher = {IOP Publishing},
	doi = {10.1088/0305-4470/16/18/029},
    url = {https://doi.org/10.1088/0305-4470/16/18/029}
}

@article{Arkin1998,
	author = {Arkin, Adam and Ross, John and McAdams, Harley H.},
	title = {\emph{Stochastic Kinetic Analysis of Developmental Pathway Bifurcation in Phage {$\lambda$}-Infected Escherichia coli Cells}},
	journal = {Genetics},
	volume = {149},
	number = {4},
	pages = {1633--1648},
	year = {1998},
	month = aug,
	issn = {1943-2631},
	publisher = {Oxford Academic},
	doi = {10.1093/genetics/149.4.1633},
    url = {https://doi.org/10.1093/genetics/149.4.1633}
}

@book{Wasshuber,
	author = {Wasshuber, Christoph},
	title = {\emph{Computational Single-Electronics}},
	journal = {SpringerLink},
	isbn = {978-3-211-83558-6},
	publisher = {Springer Vienna},
	address = {Wien, Austria},
    year={2001},
	url={https://link.springer.com/book/9783211835586}
}

@article{Moreira2023,
	author = {Moreira, Saulo V. and Samuelsson, Peter and Potts, Patrick P.},
	title = {\emph{Stochastic Thermodynamics of a Quantum Dot Coupled to a Finite-Size Reservoir}},
	journal = {Phys. Rev. Lett.},
	volume = {131},
	number = {22},
	pages = {220405},
	year = {2023},
	month = dec,
	issn = {1079-7114},
	publisher = {American Physical Society},
	doi = {10.1103/PhysRevLett.131.220405},
    url = {https://doi.org/10.1103/PhysRevLett.131.220405}
}

@article{hanggi1982,
	author = {Hanggi, Peter},
	title = {\emph{Nonlinear fluctuations: The problem of deterministic limit and reconstruction of stochastic dynamics}},
	journal = {Phys. Rev. A},
	volume = {25},
	number = {2},
	pages = {1130--1136},
	year = {1982},
	month = feb,
	issn = {2469-9934},
	publisher = {American Physical Society},
	doi = {10.1103/PhysRevA.25.1130},
    url = {https://doi.org/10.1103/PhysRevA.25.1130}
}

@article{Bressloff2017,
	author = {Bressloff, Paul C.},
	title = {\emph{Stochastic switching in biology: from genotype to phenotype}},
	journal = {J. Phys. A: Math. Theor.},
	volume = {50},
	number = {13},
	pages = {133001},
	year = {2017},
	month = feb,
	issn = {1751-8121},
	publisher = {IOP Publishing},
	doi = {10.1088/1751-8121/aa5db4},
    url = {https://doi.org/10.1088/1751-8121/aa5db4}
}

@article{hanggi1984,
	author = {Hänggi, Peter and Grabert, Hermann and Talkner, Peter and Thomas, Harry},
	title = {\emph{Bistable systems: Master equation versus Fokker-Planck modeling}},
	journal = {Phys. Rev. A},
	volume = {29},
	number = {1},
	pages = {371--378},
	year = {1984},
	month = jan,
	issn = {2469-9934},
	publisher = {American Physical Society},
	doi = {10.1103/PhysRevA.29.371},
    url = {https://doi.org/10.1103/PhysRevA.29.371}
}

@article{Gao2021,
	author = {Gao, Chloe Ya and Limmer, David T.},
	title = {\emph{Principles of low dissipation computing from a stochastic circuit model}},
	journal = {Phys. Rev. Res.},
	volume = {3},
	number = {3},
	pages = {033169},
	year = {2021},
	month = aug,
	issn = {2643-1564},
	publisher = {American Physical Society},
	doi = {10.1103/PhysRevResearch.3.033169},
    url = {https://doi.org/10.1103/PhysRevResearch.3.033169}
}

@article{Lazarescu2019,
	author = {Lazarescu, Alexandre and Cossetto, Tommaso and Falasco, Gianmaria and Esposito, Massimiliano},
	title = {\emph{Large deviations and dynamical phase transitions in stochastic chemical networks}},
	journal = {J. Chem. Phys.},
	volume = {151},
	number = {6},
	pages = {064117},
	year = {2019},
	month = aug,
	issn = {0021-9606},
	publisher = {AIP Publishing},
	doi = {10.1063/1.5111110},
    url = {https://doi.org/10.1063/1.5111110}
}

@article{Schmiedl2007,
	author = {Schmiedl, Tim and Seifert, Udo},
	title = {\emph{Stochastic thermodynamics of chemical reaction networks}},
	journal = {J. Chem. Phys.},
	volume = {126},
	number = {4},
	pages = {044101},
	year = {2007},
	month = jan,
	issn = {0021-9606},
	publisher = {AIP Publishing},
	doi = {10.1063/1.2428297},
    url = {https://doi.org/10.1063/1.2428297}
}

@article{rezaei2020,
	author = {Rezaei, Elahe and Donato, Marco and Patterson, William R. and Zaslavsky, Alexander and Bahar, R. Iris},
	title = {\emph{Fundamental Thermal Limits on Data Retention in Low-Voltage CMOS Latches and SRAM}},
	journal = {IEEE Trans. Device Mater. Rel.},
	volume = {20},
	number = {3},
	pages = {488--497},
	year = {2020},
	month = may,
	publisher = {IEEE},
	doi = {10.1109/TDMR.2020.2996627},
    url = {https://doi.org/10.1109/TDMR.2020.2996627}
}

@article{gillespie2976general,
    author = {Gillespie, Daniel T.},
    doi = {10.1016/0021-9991(76)90041-3},
    issn = {0021-9991},
    journal = {J. Comput. Phys.},
    number = {4},
    pages = {403-434},
    title = {\emph{A general method for numerically simulating the stochastic time evolution of coupled chemical reactions}},
    url = {https://doi.org/10.1016/0021-9991(76)90041-3},
    volume = {22},
    year = {1976}}

@article{Horowitz2015,
	author = {Horowitz, Jordan M.},
	title = {\emph{Diffusion approximations to the chemical master equation only have a consistent stochastic thermodynamics at chemical equilibrium}},
	journal = {J. Chem. Phys.},
	volume = {143},
	number = {4},
	year = {2015},
	month = jul,
	issn = {0021-9606},
	publisher = {AIP Publishing},
    url = {https://doi.org/10.1063/1.4927395}
}

@book{Gardiner,
	author = {Gardiner, Crispin},
	title = {\emph{Stochastic Methods}},
	journal = {SpringerLink},
	isbn = {978-3-540-70712-7},
	publisher = {Springer},
	address = {Berlin, Germany},
    year={2009},
	url = {https://link.springer.com/book/9783540707127}
}

@article{santolin2024,
  title = {\emph{Bridging Freidlin-Wentzell large deviations theory and stochastic thermodynamics}},
  author = {Santolin, Davide and Freitas, Nahuel and Esposito, Massimiliano and Falasco, Gianmaria},
  journal = {Phys. Rev. E},
  volume = {111},
  issue = {2},
  pages = {024106},
  numpages = {12},
  year = {2025},
  month = {Feb},
  publisher = {American Physical Society},
  doi = {10.1103/PhysRevE.111.024106},
  url = {https://link.aps.org/doi/10.1103/PhysRevE.111.024106}
}

@article{falasco2023,
  title = {\emph{Macroscopic stochastic thermodynamics}},
  author = {Falasco, Gianmaria and Esposito, Massimiliano},
  journal = {Rev. Mod. Phys.},
  volume = {97},
  issue = {1},
  pages = {015002},
  numpages = {46},
  year = {2025},
  month = {Jan},
  publisher = {American Physical Society},
  doi = {10.1103/RevModPhys.97.015002},
  url = {https://link.aps.org/doi/10.1103/RevModPhys.97.015002}
}

@article{vellela2009,
  title={\emph{Stochastic dynamics and non-equilibrium thermodynamics of a bistable chemical system: the {Schl\"ogl} model revisited}},
  author={Vellela, Melissa and Qian, Hong},
  journal={J. R. Soc. Interface},
  volume={6},
  number={39},
  pages={925--940},
  year={2009},
  publisher={The Royal Society},
  doi={10.1098/rsif.2008.0476},
}

@article{VanVu2023,
	author = {Van Vu, Tan and Saito, Keiji},
	title = {\emph{Thermodynamic Unification of Optimal Transport: Thermodynamic Uncertainty Relation, Minimum Dissipation, and Thermodynamic Speed Limits}},
	journal = {Phys. Rev. X},
	volume = {13},
	number = {1},
	pages = {011013},
	year = {2023},
	month = feb,
	issn = {2160-3308},
	publisher = {American Physical Society},
	doi = {10.1103/PhysRevX.13.011013},
    url = {https://doi.org/10.1103/PhysRevX.13.011013}
}

@article{VanVu2024,
  title = {\emph{Dissipation, quantum coherence, and asymmetry of finite-time cross-correlations}},
  author = {Van Vu, Tan and Vo, Van Tuan and Saito, Keiji},
  journal = {Phys. Rev. Res.},
  volume = {6},
  issue = {1},
  pages = {013273},
  numpages = {18},
  year = {2024},
  month = {Mar},
  publisher = {American Physical Society},
  doi = {10.1103/PhysRevResearch.6.013273},
  url = {https://link.aps.org/doi/10.1103/PhysRevResearch.6.013273}
}

@article{kubo1973fluctuation,
  title={\emph{Fluctuation and relaxation of macrovariables}},
  author={Kubo, Ryogo and Matsuo, Kazuhiro and Kitahara, Kazuo},
  journal={J. Stat. Phys.},
  volume={9},
  pages={51--96},
  year={1973},
  publisher={Springer},
  doi={10.1007/BF01016862},
  url={https://doi.org/10.1007/BF01016862}
}

@book{risken,
	author = {Risken, Hannes},
	title = {\emph{The Fokker-Planck Equation: Methods of Solution and Applications}},
    series = {Springer Series in Synergetics},
    editor = {Hermann Haken},
    volume = {18},
    year = {1989},
    edition = {Second},
	isbn = {978-3-540-61530-9},
	publisher = {Springer-Verlag},
	address = {Berlin},
	url = {https://doi.org/10.1007/978-3-642-61544-3}
}

@book{vankampen,
	author = {van Kampen, N. G.},
	title = {\emph{Stochastic Processes in Physics and Chemistry}},
    year = {2007},
    edition = {Third},
	isbn = {978-0-444-52965-7},
	publisher = {Elsevier},
	address = {Amsterdam}
}

@book{reichl,
	author = {Reichl, L.~E.},
	title = {\emph{A Modern Course in Statistical Physics}},
    year = {1998},
    edition = {Second},
	isbn = {0-471-59520-9},
	publisher = {John Wiley \& Sons, Inc.},
	address = {New York}
}

@ARTICLE{hanggi1988bistability,
  author={Hänggi, Peter and Jung, Peter},
  journal={IBM J. Res. Dev.}, 
  title={\emph{Bistability in active circuits: Application of a novel {Fokker}-{Planck} approach}}, 
  year={1988},
  volume={32},
  number={1},
  pages={119-126},
  doi={10.1147/rd.321.0119},
  url={https://doi.org/10.1147/rd.321.0119}
}

@article{gillespie1977exact,
	author = {Gillespie, Daniel T. },
	date = {1977/12/01},
	doi = {10.1021/j100540a008},
	isbn = {0022-3654},
	journal = {J. Phys. Chem.},
	journal1 = {The Journal of Physical Chemistry},
	journal2 = {J. Phys. Chem.},
	month = {12},
	number = {25},
	pages = {2340--2361},
	publisher = {American Chemical Society},
	title = {\emph{Exact stochastic simulation of coupled chemical reactions}},
	type = {doi: 10.1021/j100540a008},
	url = {https://doi.org/10.1021/j100540a008},
	volume = {81},
	year = {1977},
	year1 = {1977}
}

@article{gillespie2000chemical,
	author = {Gillespie, Daniel T.},
	doi = {10.1063/1.481811},
	issn = {0021-9606},
	journal = {J. Chem. Phys.},
	month = {07},
	number = {1},
	pages = {297-306},
	title = {\emph{The chemical Langevin equation}},
	url = {https://doi.org/10.1063/1.481811},
	volume = {113},
	year = {2000}}

@article{freitas2022emergent,
  title={\emph{Emergent second law for non-equilibrium steady states}},
  author={Freitas, Jos{\'e} Nahuel and Esposito, Massimiliano},
  journal={Nat. Commun.},
  volume={13},
  number={1},
  pages={5084},
  year={2022},
  publisher={Nature Publishing Group UK London},
  doi={10.1038/s41467-022-32700-7},
}

@article{freitas2021linear,
  title={\emph{Linear response in large deviations theory: a method to compute non-equilibrium distributions}},
  author={Freitas, Nahuel and Falasco, Gianmaria and Esposito, Massimiliano},
  journal={New J. Phys.},
  volume={23},
  number={9},
  pages={093003},
  year={2021},
  publisher={IOP Publishing},
  doi={10.1088/1367-2630/ac1bf5},
}

@article{freitasStochastic2021,
  title = {\emph{Stochastic Thermodynamics of Nonlinear Electronic Circuits: A Realistic Framework for Computing Around $kT$}},
  author = {Freitas, Nahuel and Delvenne, Jean-Charles and Esposito, Massimiliano},
  journal = {Phys. Rev. X},
  volume = {11},
  issue = {3},
  pages = {031064},
  numpages = {27},
  year = {2021},
  month = {Sep},
  publisher = {American Physical Society},
  doi = {10.1103/PhysRevX.11.031064},
  url = {https://link.aps.org/doi/10.1103/PhysRevX.11.031064}
}

@article{freitasReliability2022,
  title = {\emph{Reliability and entropy production in nonequilibrium electronic memories}},
  author = {Freitas, Nahuel and Proesmans, Karel and Esposito, Massimiliano},
  journal = {Phys. Rev. E},
  volume = {105},
  issue = {3},
  pages = {034107},
  numpages = {9},
  year = {2022},
  month = {Mar},
  publisher = {American Physical Society},
  doi = {10.1103/PhysRevE.105.034107},
  url = {https://link.aps.org/doi/10.1103/PhysRevE.105.034107}
}

@article{bouchetGeneralization2016,
	author = {Bouchet, Freddy and Reygner, Julien},
	date = {2016/12/01},
	doi = {10.1007/s00023-016-0507-4},
	id = {Bouchet2016},
	isbn = {1424-0661},
	journal = {Ann. Henri Poincar{\'e}},
	number = {12},
	pages = {3499--3532},
	title = {\emph{Generalisation of the Eyring--Kramers Transition Rate Formula to Irreversible Diffusion Processes}},
	url = {https://doi.org/10.1007/s00023-016-0507-4},
	volume = {17},
	year = {2016}}

@article{chang1970practical,
	author = {Chang, J.S. and Cooper, G.},
	doi = {10.1016/0021-9991(70)90001-X},
	issn = {0021-9991},
	journal = {J. Comput. Phys.},
	number = {1},
	pages = {1-16},
	title = {\emph{A practical difference scheme for {Fokker-Planck} equations}},
	url = {https://doi.org/10.1016/0021-9991(70)90001-X},
	volume = {6},
	year = {1970}
}

@article{mckane2005predator,
  title={\emph{Predator-prey cycles from resonant amplification of demographic stochasticity}},
  author={McKane, Alan J and Newman, Timothy J},
  journal={Phys. Rev. Lett.},
  volume={94},
  number={21},
  pages={218102},
  year={2005},
  publisher={APS},
  doi={10.1103/PhysRevLett.94.218102},
  url={https://doi.org/10.1103/PhysRevLett.94.218102}
}

@article{fisher2014transition,
  title={\emph{The transition between the niche and neutral regimes in ecology}},
  author={Fisher, Charles K and Mehta, Pankaj},
  journal={Proc. Natl. Acad. Sci. U.S.A.},
  volume={111},
  number={36},
  pages={13111--13116},
  year={2014},
  publisher={National Acad Sciences},
  url={https://doi.org/10.1073/pnas.1405637111}
}

@article{reichenbach2006coexistence,
  title={\emph{Coexistence versus extinction in the stochastic cyclic Lotka-Volterra model}},
  author={Reichenbach, Tobias and Mobilia, Mauro and Frey, Erwin},
  journal={Phys. Rev. E},
  volume={74},
  number={5},
  pages={051907},
  year={2006},
  publisher={APS},
  doi={10.1103/PhysRevE.74.051907},
  url={https://doi.org/10.1103/PhysRevE.74.051907}
}

@article{ovaskainen2010stochastic,
  title={\emph{Stochastic models of population extinction}},
  author={Ovaskainen, Otso and Meerson, Baruch},
  journal={Trends Ecol. Evol.},
  volume={25},
  number={11},
  pages={643--652},
  year={2010},
  publisher={Elsevier},
  doi={10.1016/j.tree.2010.07.001},
  url={https://doi.org/10.1016/j.tree.2010.07.001}
}

\newpage

\onecolumngrid 
\appendix

\pagebreak
\widetext
\begin{center}
\textbf{\large Supplemental Materials: Improved diffusive approximation of Markov jump processes close to equilibrium}

\vspace{10pt} 

David Roberts, Trevor McCourt, Geremia Massarelli, Jeremy Rothschild, and Nahuel Freitas

\vspace{10pt} 

\textit{Extropic Corporation, Cambridge, Massachusetts, USA}
\end{center}

\section{First-order analysis of the stationary large-deviation function}
\label{app:linear-response}

In this appendix, we demonstrate that the stationary large-deviation function of our improved diffusive approximation, $I_\text{IMP}(\bm{x})$, agrees with the exact large-deviation function of the underlying Markov jump process, $I(\bm{x})$, up to first order in the parameter $\epsilon$ that quantifies the departure from detailed balance.

As established in Sect.~III in the main text, both rate functions agree at equilibrium, i.e., their zeroth-order terms are identical: $I_\text{IMP}^{(0)}(\bm{x}) = I^{(0)}(\bm{x}) = \phi(\bm{x})$. We now seek to prove that their first-order corrections, $I_\text{IMP}^{(1)}(\bm{x})$ and $I^{(1)}(\bm{x})$, are also identical. Our analysis relies on the assumption that the deterministic dynamics at equilibrium, governed by the drift $\bm{u}^{(0)}(\bm{x})$, has fixed-point attractors whose basins of attraction partition the entire state space.

We begin by expanding the macroscopic drift $\bm{u}(\bm{x})$ and diffusion $\bm{d}(\bm{x})$ tensors to first order in $\epsilon$:
\begin{align}
    \bm{u}(\bm{x}) &= \bm{u}^{(0)}(\bm{x}) + \epsilon \bm{u}^{(1)}(\bm{x}) + O(\epsilon^2) \\
    \bm{d}(\bm{x}) &= \bm{d}^{(0)}(\bm{x}) + \epsilon \bm{d}^{(1)}(\bm{x}) + O(\epsilon^2)
\end{align}
Substituting these expansions, along with $I_\text{IMP}(\bm{x}) = \phi(\bm{x}) + \epsilon I^{(1)}_\text{IMP}(\bm{x}) + O(\epsilon^2)$, into the Hamilton-Jacobi equation for the rate function, given by Eq.~\eqref{eq:FP-rate-function} in the main text, and collecting the first-order terms in $\epsilon$ yields:
\begin{align}
    \bm{u}^{(0)}(\bm{x})\cdot \nabla I^{(1)}_\text{IMP}(\bm{x}) = -\bm{u}^{(1)}(\bm{x})\cdot \nabla \phi(\bm{x}) - \nabla \phi(\bm{x}) \cdot \bm{d}^{(1)}(\bm{x})\cdot \nabla \phi(\bm{x}).
\end{align}
Using the equilibrium Einstein relation, $\bm{u}^{(0)}(\bm{x}) = -\bm{d}^{(0)}(\bm{x})\cdot \nabla \phi(\bm{x})$, we can rewrite the right-hand side to find:
\begin{align}
    \bm{u}^{(0)}(\bm{x})\cdot \nabla I^{(1)}_\text{IMP}(\bm{x}) &= \lim_{\epsilon\to 0} \frac{1}{\epsilon} \left( \bm{u}(\bm{x}) + \bm{d}(\bm{x})\cdot \nabla \phi(\bm{x}) \right) \cdot \left(-\nabla \phi(\bm{x}) \right) \nonumber \\
    &= -\sum_{\rho>0}\bigg(\omega_\rho^{(0)}(\bm{x})-\omega_{-\rho}^{(0)}(\bm{x})\bigg)w_\rho(\bm{x}), \label{eq:the_rate}
\end{align}
where $\omega_\rho^{(0)}(\bm{x}) \equiv \lim_{\epsilon\to 0} \omega_\rho(\bm{x})$ denotes the macroscopic jump rate at detailed balance.

Crucially, this result is identical to the known first-order equation for the exact rate function $I(\bm{x})$ of the Markov jump process, as derived in Ref.~\cite{freitas2021linear} (cf. Eq.~(11) therein):
\begin{align}
    \bm{u}^{(0)}(\bm{x})\cdot\nabla I^{(1)}(\bm{x}) = -\sum_{\rho>0}\bigg(\omega_\rho^{(0)}(\bm{x})-\omega_{-\rho}^{(0)}(\bm{x})\bigg)w_\rho(\bm{x}). \label{eq:the_rate2}
\end{align}
Comparing Eq.~\eqref{eq:the_rate} and Eq.~\eqref{eq:the_rate2}, we see that the gradients of the first-order corrections are related by $\bm{u}^{(0)}(\bm{x}) \cdot \nabla I^{(1)}(\bm{x}) = \bm{u}^{(0)}(\bm{x}) \cdot \nabla I^{(1)}_\text{IMP}(\bm{x})$. Since the zeroth-order terms also agree, this implies that the first-order expansions of the total rate functions also satisfy this relation:
\begin{equation}
    \bm{u}^{(0)}(\bm{x})\cdot \nabla I^{(\leq 1)}(\bm{x}) =\bm{u}^{(0)}(\bm{x})\cdot \nabla I^{(\leq 1)}_\text{IMP}(\bm{x}),
\end{equation}
where we have used $f^{(\leq k)}\equiv \sum_{j\leq k}f^{(j)}\epsilon^j$ to denote the $k$th-order approximant of any function $f$ of $\epsilon$. To show that the functions themselves are equal, consider their difference, $F(\bm{x})\equiv I^{(\leq 1)}(\bm{x}) - I_\text{IMP}^{(\leq 1)}(\bm{x})$. The directional derivative of $F(\bm{x})$ along any trajectory $\bm{x}^{(0)}_\tau$ of the equilibrium dynamics, $\dot{\bm{x}}^{(0)}_\tau = \bm{u}^{(0)}(\bm{x}^{(0)}_\tau)$, is zero:
\begin{equation}
    \frac{d}{d\tau}F(\bm{x}^{(0)}_\tau) = \dot{\bm{x}}^{(0)}_\tau \cdot \nabla F(\bm{x}^{(0)}_\tau) = \bm{u}^{(0)}(\bm{x}^{(0)}_\tau)\cdot \nabla F(\bm{x}^{(0)}_\tau) = 0.
\end{equation}
This means that $F(\bm{x})$ is constant along any such trajectory. Assuming that every point $\bm{x}$ belongs to a basin of attraction $\mathcal{B}_j$ of a stable fixed point $\bm{x}_j^*$, the trajectory starting from $\bm{x}$ will flow to $\bm{x}_j^*$. Therefore, $F(\bm{x}) = F(\bm{x}_j^*)$, which implies that $F(\bm{x})$ is constant throughout each basin of attraction. If we further assume that $F(\bm{x})$ is continuous across the boundaries of these basins, it must be a global constant. Since large-deviation functions are defined only up to an additive constant, we are free to set this constant difference to zero.

Thus, we conclude that $I^{(\leq 1)}(\bm{x}) = I^{(\leq 1)}_\text{IMP}(\bm{x})$, confirming that our improved diffusion approximation correctly captures the stationary fluctuations to first order in the departure from equilibrium.

\begin{figure}
    \centering
    \includegraphics[width=0.7\linewidth]{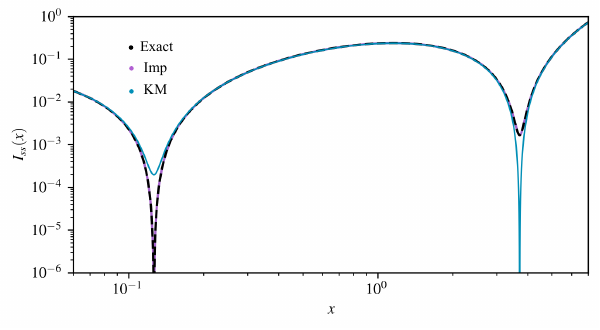}
    \caption{{\bf Quasipotential of the Schl\"ogl model.} The steady-state self-information $I_\text{ss}(x)$ for the Schl\"ogl model in its bistable regime. We compare the exact solution with the improved (Imp) and Kramers-Moyal (KM) diffusion approximations. For this one-dimensional system, the improved approximation is exact. The Kramers-Moyal approximation predicts the wrong global minimum, erroneously suggesting the right-most state is metastable and that the left-most state is stable. Here, $k_{+1}[A]=3, k_{+2}[B] = 1.305/4$, $k_{-1}= 0.6$, $k_{-2} = 2.95$ (cf. Eq. \eqref{eq:my_reaction}).}
    \label{fig:schlogl}
\end{figure}

\section{Keizer's paradox \& the Schl\"ogl model}
In this appendix, we consider the Schlögl model, which is an example of a Markov jump process for which the Kramers-Moyal approximation is known to yield the incorrect macroscopic limit in certain parameter regimes \cite{vellela2009}. It describes a single chemical species, X, whose count changes via two reversible reactions where the concentrations of species A and B are held constant:
\begin{align}
    A + 2X \xrightleftharpoons[k_{+1}]{k_{-1}} 3X, \quad B \xrightleftharpoons[k_{+2}]{k_{-2}} X.\label{eq:my_reaction}
\end{align}
The system's volume, $\Omega$, acts as the macroscopic scaling parameter. For a state with concentration $x$, the law of mass action yields the following macroscopic rates for transitions that increase or decrease the number of X molecules:
\begin{align}
    \omega_+(x) = k_{+1}[A] x^2 + k_{+2}[B], \quad \omega_-(x) = k_{-1} x^3 + k_{-2}x.
\end{align}
As a one-dimensional, nearest-neighbor jump process, the model's steady-state distribution is exactly solvable. This makes it an ideal testbed for comparing the performance of different diffusion approximations.

In Figure~\ref{fig:schlogl}, we plot the steady-state quasipotential, $I_\text{ss}(x)$, for a choice of parameters that places the system within its bistable regime. As predicted by the theory for 1D systems, the improved diffusion approximation exactly reproduces the quasipotential of the original Markov jump process. However, as noted in \cite{vellela2009}, the KM approximation incorrectly identifies the globally stable state of the system. This example highlights how the standard KM truncation can fail to capture even the most fundamental features of a system's steady state, a deficiency that the improved diffusion approximation corrects.

\section{Scaling expansion}
\label{ap:expansion}

In this appendix, given that the function \(F_\rho (\bx)\) admits the expansion of Eq.~(11) in the main text, we show that the leading-order contribution is given by \(f_\rho(\bm{x})\) in Eq.~(20) in the main text. The functions \(F_\rho (\bx)\) and \(f_\rho(\bm{x})\) are, respectively, given by the infinite sums in Eqs.~(11) and (20) in the main text.
Let us denote the individual terms of the sums as
\begin{gather}
    F_{\rho,k}(\bm{x}) \coloneqq
    e^{\mathcal{I}(\bm{x},t)} \frac{1}{k!} \left(\frac{-\bm{\Delta}_\rho \cdot \nabla}{\Omega}\right)^{k-2}\left[\lambda_\rho(\bm{x}) e^{-\mathcal{I}(\bm{x},t)}\right], \label{eq:f-infinite-sum}
    \\
    f_{\rho,k}(\bm{x}) \coloneqq 
    \omega_\rho(\bm{x}) 
    \frac{1}{k!} (\bm{\Delta}_\rho \cdot \nabla I(\bm{x},t))^{k-2}, \label{eq:f0-infinite-sum}
\end{gather}
where \( k \in \{2, 3, 4, \dots\} \).
Thus, the functions \(F_\rho (\bx)\) and \(f_{\rho}(\bm{x})\) are given by
\begin{gather}
    F_{\rho}(\bm{x}) = \sum_{k=2}^\infty F_{\rho,k}(\bm{x}),
    \\
    f_{\rho}(\bm{x}) = \sum_{k=2}^\infty f_{\rho,k}(\bm{x}).
\end{gather}
We will show by mathematical induction that \(F_{\rho,k} (\bx) = \Omega f_{\rho,k}(\bm{x}) + \mathcal{O}(\Omega^0)\) for all \( k \geq 2 \); this is our induction hypothesis. 
This, in turn, implies that the infinite sums are related by \(F_{\rho} (\bx) = \Omega f_{\rho}(\bm{x}) + \mathcal{O}(\Omega^0)\), as required.

\paragraph{Base step}
Consider \(k = 2\) as the base step: we must show that \(F_{\rho,2} (\bx) = \Omega f_{\rho,2}(\bm{x}) + \mathcal{O}(\Omega^0)\).
Starting with the expression in Eq.~\eqref{eq:f-infinite-sum}, we have
\begin{subequations}
\begin{align}
    F_{\rho,2}(\bm{x}) 
    &= \frac{1}{2} \lambda_\rho(\bm{x})
    \\
    &= \Omega \frac{1}{2} \omega_\rho(\bm{x}) + \mathcal{O}(\Omega^0)
    \\
    &= \Omega f_{\rho,2}(\bm{x}) + \mathcal{O}(\Omega^0).
\end{align}
\end{subequations}
In the last line, we have recognized the form of Eq.~\eqref{eq:f0-infinite-sum}, as required.

\paragraph{Induction step}
Assuming that \(F_{\rho,k} (\bx) = \Omega f_{\rho,k}(\bm{x}) + \mathcal{O}(\Omega^0)\), we prove that \(F_{\rho,k+1} (\bx) = \Omega f_{\rho,k+1}(\bm{x}) + \mathcal{O}(\Omega^0)\).
Rearranging terms, we find that
\begin{subequations}
\begin{equation}
    F_{\rho,k+1}(\bx) = e^{\mathcal{I}(\bm{x},t)} \frac{1}{k+1} 
    \left(\frac{-\bm{\Delta}_\rho \cdot \nabla}{\Omega}\right) \left[ e^{-\mathcal{I}(\bm{x},t)} F_{\rho,k}(\bx) \right].
\end{equation}
Then, applying the induction hypothesis, differentiating, and using the expansion for the self-information in Eq.~(18) in the main text, we arrive at
\begin{align}
    F_{\rho,k+1}(\bx)
    &= \Omega \frac{\bm{\Delta}_\rho \cdot \nabla I(\bm{x},t)}{k+1} f_{\rho,k}(\bm{x}) + \mathcal{O}(\Omega^0)
    \\
    &= \Omega f_{\rho,k+1}(\bm{x}) + \mathcal{O}(\Omega^0),
\end{align}
\end{subequations}
as required.

\section{Discretization and numerical solution of the Fokker-Planck equation}

In the present work, we carried out large-scale numerical solutions of the Fokker-Planck equation in order to assess the performance of the improved diffusion approximation vis-à-vis the standard diffusion approximation for certain MJP systems.
We now outline our general procedure for numerically solving a Fokker-Planck equation,
\begin{equation} \label{eq:appendix_fokker_planck_equation}
    \partial_t P(\bx,t)
    \ =\ -\partial_n \Bigl( \mu_n(\bm{x}) P(\bm{x},t)-\partial_m \bigl(D_{nm}(\bm{x}) P(\bm{x},t)\bigr) \Bigr)
    \ =\ \mathrm{L}^\text{FP}(\bx) P(\bx,t),
\end{equation}
where, in the last step, we have introduced the linear differential operator \(\mathrm{L}^\text{FP}(\bx)\).
The main objective of this appendix is to justify our choice of \(\mathrm{L}^\text{FP}_{ij}\), the discretized version of \(\mathrm{L}^\text{FP}(\bx)\), for it is this object that we diagonalize numerically in order to find the approximate spectrum of \(\mathrm{L}^\text{FP}(\bx)\).

It is often advantageous to employ discretization schemes that \emph{exactly} preserve certain key features of the equation at the level of the approximate solution, such as conservation of probability and positivity of solutions, which are central to the interpretation of \(P(\bx, t)\) as a probability distribution.
Although any generic discretization scheme should at least \emph{approximately} reproduce these features, this may come at the cost of using very fine discretizations.

The first and most well-known work on this topic is perhaps that of Chang and Cooper~\cite{chang1970practical}, in which they address the issue of conservation of probability by adequately differencing the equation and by using reflecting boundary conditions.
Furthermore, they also introduce an “exponential upwind” scheme that guarantees the positivity of solutions.

In the present work, we partly borrow from them, using a scheme that conserves probability exactly.
In contrast, as we did not observe any issues regarding the positivity of solutions, we did not introduce additional mathematical complexity to guarantee this property.

\subsection{Conservation of probability in the bulk} \label{sect:probability_conservation_bulk}

A priori, one may be inclined to rewrite the equation using the product rule, obtaining
\begin{equation} \label{eq:appendix_fokker_planck_product_rule}
    \mathrm{L}^\text{FP}(\bx) P(\bx,t) =
    \Bigl(-\partial_n \mu_n(\bx) + \partial_n\partial_m D_{nm}(\bx) \Bigr) P(\bx, t)
    + \Bigl( - \mu_n(\bx) + 2 \partial_m D_{nm}(\bx) \Bigr) \partial_n P(\bx, t)
    + D_{nm}(\bx) \partial_n \partial_m P(\bx, t).
\end{equation}
Then, the first and second derivatives of \(P(\bx, t)\) would be approximated using finite differences, while those of \(\mu_n(\bx)\) and \(D_{nm}(\bx)\) can be calculated analytically.
This scheme, however, has the significant drawback of not conserving probability: every point in the (discretized) space may act as a source or sink~\cite{chang1970practical}.
Let us see why.

At the level of the partial differential equation, conservation of probability arises because Eq.~\eqref{eq:appendix_fokker_planck_equation} has the form of a continuity equation.
Explicitly, one can write
\begin{equation} \label{eq:prob_conservation_noboundary}
    \frac{\mathrm{d}}{\mathrm{d}t} \int \mathrm{d}^d x\ P(\bx, t) = \int \mathrm{d}^d x\ \partial_t P(\bx, t) = - \int \mathrm{d}^d x\ \partial_n J_n(\bx, t) = 0,
\end{equation}
where \( J_n(\bx, t) \coloneqq \mu_n(\bm{x}) P(\bm{x},t) - \partial_m \bigl( D_{nm}(\bm{x}) P(\bm{x},t) \bigr) \).
In the last step, we recognized that the integral of a divergence vanishes since, for now, we are ignoring boundary contributions.
Hence, probability is conserved.

After discretizing space, we may write the Fokker-Planck equation as
\(
    \frac{\mathrm{d}P_i(t)}{\mathrm{d}t} = \sum_j \mathrm{L}^\text{FP}_{ij} P_j(t),
\)
where \(\mathrm{L}^\text{FP}_{ij}\) is the matrix encoding the finite-difference operator, a correspondence which we denote \(\mathrm{L}^\text{FP} \rightarrow \mathrm{L}^\text{FP}_{ij}\).
Now, conservation of probability takes the form
\begin{equation}
    \frac{\mathrm{d}}{\mathrm{d}t} \sum_i P_i(t) = \sum_j \left( \sum_i \mathrm{L}^\text{FP}_{ij} \right) P_j(t) \overset{!}{=} 0,
\end{equation}
that is, the \emph{columns} of \(\mathrm{L}^\text{FP}_{ij}\) must sum to zero in order for probability to be conserved.

It is now apparent why the discretized version of 
Eq.~\eqref{eq:appendix_fokker_planck_product_rule} does not conserve probability: take, for example, the term \(\mathrm{L} = \mu_1(\bx) \partial_1\), which enters into \(\mathrm{L}^\text{FP}\).
Denoting the discretized version of \(\partial_1\) as \(\partial_1 \rightarrow \Delta_{ij}\), we may write
\begin{equation}
    \mathrm{L} = \mu_1(\bx) \partial_1
    \quad\rightarrow\quad
    \mathrm{L}_{ij} = \sum_k M_{ik} \Delta_{kj},
\end{equation}
where \(M_{ik} = \delta_{ik} \mu_1(\bx_i)\).
Although \(\Delta_{ij}\) satisfies
\begin{equation} \label{eq:deriv_columns_sum_to_zero}
    \sum_k \Delta_{kj} = \sum_j \Delta_{kj} = 0 \quad \text{(away from any boundaries)},
\end{equation}
one generally finds that probability is not exactly conserved in the approximate solution:
\begin{equation}
    \sum_i \mathrm{L}_{ij} = \sum_{i,k} M_{ik} \Delta_{kj} = \sum_i \mu_1(\bx_i) \Delta_{ij} \neq 0.
\end{equation}
If, instead, one refrains from using the product rule and directly differences the equation (including the coefficients), one ends up with the operator
\begin{equation}
    \mathrm{L}' = \partial_1 \mu_1(\bx)
    \quad\rightarrow\quad
    \mathrm{L}'_{ij} = \sum_k \Delta_{ik} M_{kj},
\end{equation}
which \emph{does} exactly conserve probability (away from any boundaries):
\begin{equation}
    \sum_i \mathrm{L}'_{ij} = \sum_{i,k} \Delta_{ik} M_{kj} = \sum_{k} \left(\sum_i \Delta_{ik}\right) M_{kj} = 0.
\end{equation}

Hence, the preferred approach is to difference Eq.~\eqref{eq:appendix_fokker_planck_equation} \emph{as is}.
Since the finite-difference approximations of the derivatives \(\partial_n\) and \(\partial_n\partial_m\) all satisfy conditions like that of Eq.\eqref{eq:deriv_columns_sum_to_zero}, the columns of the matrix operator \(\mathrm{L}^\text{FP}_{ij}\) will sum to zero (at least, away from boundary points).

\subsection{Conservation of probability at the boundaries: reflecting boundary conditions}

In Eq.~\eqref{eq:prob_conservation_noboundary}, we ignored the effect of the boundary.
Including its effect, one finds, instead, 
\begin{equation}
    \frac{\mathrm{d}}{\mathrm{d}t} \int_\Omega \mathrm{d}^d x\ P(\bx, t) = - \oint_{\partial\Omega} \mathrm{d}S_n \ J_n(\bx, t),
\end{equation}
where \(\Omega\) denotes the region of integration and \(\partial\Omega\) denotes its boundary.

Reflecting boundary conditions are those for which \(\mathrm{d}S_n \ J_n(\bx, t) = 0\) at boundary points, meaning that no probability escapes through the boundary.
For a boundary at infinity, reflecting boundary conditions coincide with \emph{clamped} boundary conditions (i.e., Dirichlet boundary conditions with vanishing probability)~\cite[Sect.~5.4]{risken}.
However, for a finite-size domain, reflecting and natural boundary conditions do not coincide, though the distinction tends to zero as the domain size is increased.

In our numerics, we exactly impose reflecting boundary conditions, making our discretization scheme probability conserving. With this choice, the highest-lying eigenvalue is precisely zero, up to numerical errors.

\subsubsection{Lowest order symmetric stencil}
For purposes of illustration, we start by considering the Fokker-Planck equation in one dimension, and discretize it using the lowest-order symmetric stencil:
\begin{equation} \label{eq:appendix_fokker_planck_equation_discrete_1D}
    \frac{\mathrm{d}P_i}{\mathrm{d}t} = - \frac{(\mu P)_{i+1} - (\mu P)_{i-1}}{2h} + \frac{(DP)_{i-1} -2(DP)_i + (DP)_{i+1}}{h^2}.
\end{equation}
Here, we have suppressed the time dependence of \(P\) for brevity, and we denote \(f_i \coloneqq f(x_i)\).
In the notation introduced in Sect.~\ref{sect:probability_conservation_bulk}, we have
\begin{equation}
    \mathrm{L}^\text{FP}_{ij}
    = \delta_{i-1,j} \frac{2 D_{j} + h \mu_{j}}{2h^2}
    - \delta_{ij} \frac{2 D_j}{h^2}
    + \delta_{i+1,j} \frac{2 D_{j} - h \mu_{j}}{2h^2}.
\end{equation}

\paragraph*{Clamped boundary conditions}
Say we are interested in the domain \(i \in \{0, 1, 2, \dots\}\), which, for simplicity, has only one boundary.
In order to impose clamped boundary conditions at the domain boundary, we introduce a fictitious point \(i = -1\), which is clamped at \(P_{-1}(t) = 0\) for all time.
The result is that \(P_{-1}\) drops out of Eq.~\eqref{eq:appendix_fokker_planck_equation_discrete_1D} for \(i=0\) (and, hence, all \(i\)):
\begin{equation}
    \frac{\mathrm{d}P_0}{\mathrm{d}t} = - \frac{(\mu P)_1}{2h} + \frac{-2(DP)_0 + (DP)_1}{h^2}.
\end{equation}
In other words, for clamped boundary conditions, the matrix \(\mathrm{L}^\text{FP}_{ij}\) is obtained by simple “truncation” at the boundaries.

It is also straightforward to see that probability is not conserved with these boundary conditions, since the corresponding column of \(\mathrm{L}^\text{FP}_{ij}\) does not sum to zero:
\begin{equation}
    \sum_i \mathrm{L}^\text{FP}_{i,0} = -D_0 + \frac{h}{2} \mu_0 \neq 0.
\end{equation}

\paragraph*{Reflecting boundary conditions}

We seek the discrete analogue of the probability current \(J(x, t)\), which will elucidate the question of how to implement reflecting boundary conditions.
In order to do so, we note that the discrete analogue of the relation
\begin{equation}
    \frac{\mathrm{d}}{\mathrm{d}t} \int_0^{+\infty} \mathrm{d}x\ P(x, t) = J(0, t)
\end{equation}
(where we have assumed that \(J(+\infty, t) = 0\)) is given by
\begin{subequations}
\begin{align}
    \frac{\mathrm{d}}{\mathrm{d}t} \sum_{i=0}^{+\infty} h P_i(t) &= h \sum_{i=0}^{+\infty} \frac{\mathrm{d}P_i(t)}{\mathrm{d}t}
    \\
    &= \frac{(\mu P)_{0} + (\mu P)_{-1}}{2} - \frac{(DP)_0 -(DP)_{-1}}{h}
    \\
    &\eqqcolon J_{-\frac{1}{2}},
\end{align}
\end{subequations}
where, in the last line, we have introduced \(J_{-\frac{1}{2}}\), the (discrete) probability current from \(i = -1\) to \(i = 0\).
It is also heartening that in terms of this discrete probability current, the Fokker-Planck equation can be rewritten in the form most closely analogous to a continuity equation:
\begin{equation}
    \frac{\mathrm{d}P_i(t)}{\mathrm{d}t} = - \frac{J_{i+\frac{1}{2}} - J_{i-\frac{1}{2}}}{h}.
\end{equation}

It is now clear that probability is conserved in \(i \geq 0\) if, and only if,
\begin{subequations}
    \begin{gather}
        J_{-\frac{1}{2}} = 0
        \\
        \Leftrightarrow \quad P_{-1}(t) = \frac{2D_0 \ - \ h \mu_0}{2D_{-1} + h \mu_{-1}} P_0(t).
    \end{gather}
\end{subequations}
Hence, we must, once again, introduce a fictitious point \(i = -1\), this time letting its value be determined by the relation above.
Then, by design, all columns of \(\mathrm{L}^\text{FP}_{ij}\) sum to zero and probability is conserved.

\subsubsection{General case}

In higher dimensionalities and when using a higher-order stencil, there is no longer a clear-cut, unique way of imposing reflecting boundary conditions.
In this case, we adopt the following pragmatic approach: we begin with the discretized Fokker-Planck operator \( \mathrm{L}^\text{FP}_{ij} \) with \emph{clamped} boundary conditions (which conserves probability everywhere but at the boundaries), and construct the operator \( \mathrm{L}^\text{FP}_{ij} - \delta_{ij} \sum_i \mathrm{L}^\text{FP}_{ij} \) (which conserves probability \emph{everywhere}).
Clearly, this is an implementation of reflecting boundary conditions, because (a) \( \mathrm{L}^\text{FP}_{ij} \) and \( \mathrm{L}^\text{FP}_{ij} - \delta_{ij} \sum_i \mathrm{L}^\text{FP}_{ij} \) differ only near the boundaries, and (b) \( \mathrm{L}^\text{FP}_{ij} - \delta_{ij} \sum_i \mathrm{L}^\text{FP}_{ij} \) conserves probability.

Furthermore, provided we are interested only in bulk properties (e.g., the low-lying eigenstates whose weight is concentrated far from the boundaries), as is the case in the present work, the details of how the reflecting boundary conditions are implemented are of no consequence.


\end{document}